\documentclass[12pt]{article}
\usepackage[english]{babel}
\usepackage{epsf,epsfig,graphicx} 

\newcommand{\be}{\begin{eqnarray}}

\newcommand{\ee}{\end{eqnarray}}

\newcommand{\mgdel}[1]{}

\newcommand{\lmdel}[1]{} 
\newcommand{\lton}{\mathrel{\lower.9ex 
  \hbox{$\stackrel{\displaystyle <}{\sim}$}}}
\newcommand{\gton}{\mathrel{\lower.9ex 
  \hbox{$\stackrel{\displaystyle >}{\sim}$}}}

\makeindex

\begin{document}

\title{ New Forms of QCD Matter Discovered at RHIC.
 }
\author{Miklos Gyulassy$^a$ and Larry McLerran$^b$ \\
     { \small\it $^a$Physics Department, Columbia University}
 {\small\it New York, NY USA}\\
     { \small\it  $^b$Physics Department 
POB 5000} 
{\small\it Brookhaven National Laboratory} 
{\small\it Upton, NY 11973 USA }\\
 }

\date{\underline{May 5, 2004}}
\maketitle

\abstract{We discuss two special limiting forms of QCD matter which 
may be produced 
at RHIC.  We conclude from the available empirical evidence 
that an equilibrated, but strongly coupled Quark Gluon Plasma has
been made in such collisions.  We also  
discuss the growing body of evidence that its source is a 
Color Glass Condensate.}

\section{Introduction}

Thirty years ago\cite{bm}, T.D. Lee suggested that it would be
interesting to explore new 
phenomena ``by distributing high energy or high nucleon
density over a relatively large volume."  In this way one could 
temporarily restore broken symmetries of the physical vacuum and possibly
create novel abnormal dense states of nuclear matter\cite{Lee:ma}.  W.
Greiner and collaborators pointed out that the required high densities
could be achieved via relativistic heavy ion
collisions\cite{Hofmann:by}.  Concurrently, Collins and Perry and
others\cite{Collins:1974ky} realized that the asymptotic freedom
property of Quantum Chromodynamics (QCD) 
implies the existence of an ultra-dense form of matter
with deconfined quarks and gluons, called later the  
Quark-Gluon Plasma (QGP)\cite{Shuryak:1978ij}.  In 1982 J.D.
Bjorken developed a relativistic hydrodynamic
theory\cite{Bjorken:uj} of the novel ``inside-out'' evolution of the
central low baryon density regions of ultra-relativistic heavy ion collisions.  
While many signatures of QGP formation were
proposed, he suggested that ``if the quark-gluon plasma is produced,
it will manifest itself in experimental signatures as yet unforeseen.
The system is after all a complicated relativistic fluid subject to
highly nonlinear forces. $\ldots$ If very interesting and novel phenomena
will be seen in ion-ion collisions, theory must develop the
capability to interpret them, if not to predict them''.

With this theoretical background, the 1983 DOE/NSAC Long Range Plan set in motion
plans that led to the construction of the Relativistic Heavy Ion Collider (RHIC) \cite{rhic}
to explore properties of ultra-dense matter above the deconfinement transition point.
Now, after its first three years of operation,
a  vast data base\cite{rhic_pubs}  on $p+p, D+Au,$ and $Au+Au$ at $\surd s=20-200$ AGeV 
has been harvested from RHIC. The published data are available through
22 (4 PRL) publications 
from BRAMHS\cite{brahmsexp},
 92 (15 PRL) from  PHENIX\cite{phenixexp}, 
34 (6 PRL) from PHOBOS\cite{phobosexp}, and 127 (21 PRL) 
from STAR\cite{starexp}. This body of data extends and builds upon
the knowledge gained about dense hadronic matter
measured at the SPS/CERN (publications include 108 NA49/35, 69 NA50/38, 26 CERES/NA45,
79 WA98/80, 32 na57/wa97) at $\surd s \le 20$ AGeV
The SPS heavy ion data already displayed several signatures that hinted
at the onset of QGP formation\cite{cernrpt}.  Based on the SPS data and 
theoretical predictions,
RHIC with its factor of 10 increase in the center of mass energy to 200 AGeV
was assured to create matter well above the deconfinement transition point.
The fourth year run of RHIC with Au+Au at 200 AGeV 
has just concluded with a gain of about another factor
$\sim 20$  integrated luminosity ($\sim 1400/\mu$b). These and future RHIC data
with  significantly upgraded
detectors  will provide even higher resolution measurements of the detailed properties of 
the new forms of matter discovered at RHIC.  

While there naturally remain many open questions and unsolved puzzles, 
a striking set of new phenomena
have been conclusively discovered at RHIC. Those phenomena furthermore 
can now be readily interpreted and predicted 
from significant advances in theory and from the 
empirical information gained from  the past 30 years from
the Bevalac, SIS, AGS, and SPS experiments.  

 In this report, 
we discuss\cite{Gyulassy:2004vg,McLerran:2004fg} the evidence that at least one and possibly
two new forms of QCD matter
have been discovered at RHIC.  We consider the Quark Gluon Plasma, 
which is a form of matter characterized by a thermal equilibrium density
matrix of a system of quarks and gluons.  We also consider the Color
Glass Condensate (CGC), which is a form of matter characterized by a
universal initial density matrix which describes high energy strongly
interacting particles - including nuclei.  The QGP is the incoherent
thermal limit of QCD matter at high temperatures while the CGC is the
coherent limit of QCD at high energies.  Since the QGP has to be
created at RHIC from the interaction of initial nuclear enhanced
coherent chromo electric magnetic fields, both limiting forms of QCD matter need to be
considered at RHIC. 

The first 275 published experimental papers from RHIC have 
of course only barely scratched the surface
of the new physics of these forms of matter, but the data are so 
striking and decisive that  
several strong physics conclusions can already be drawn. 
They establish empirically
that a special form of strongly coupled QGP (sQGP) 
exists with remarkable properties. 
In addition, there is growing evidence
that its source is well described by a saturated gluon CGC initial state. 
These RHIC discoveries and those at the SPS/CERN
pave a clear path for future systematic studies of these new forms
of matter in the laboratory.

We begin in section 2 by describing the basic ingredients of the QGP hypothesis.
The hypothesis concerning the existence and properties of this form of
matter has a firm basis in QCD. Much is known about its theoretical
properties on the basis of numerical computation within QCD (lattice
gauge theory).  In section 3 we review the CGC hypothesis, which is newer but
is also based firmly in QCD.  The CGC hypothesis can be tested in a
wide range of experimental environments (HERA, RHIC,LHC,eRHIC) not
restricted to heavy ion collisions.  It is newer, and because of this,
somewhat more tenuous than is the QGP hypothesis.

The scientific method is based on the paradigm that theories are
tested by falsification.  This is an important concept, since 
simple models with many adjustable parameters are often
used to ``fit'' heavy ion  data.
 However, one of the most
compelling motivations for extending studies of heavy ion physics into the
ultra-relativistic RHIC energy frontier $\surd{s}\sim 20- 200 $ AGeV, is that
 predictions, based on the QCD theory itself, of new
physics  under extreme conditions of temperature and
energy  can be tested.  For  RHIC and higher energies
controlled QCD theoretic approximations and methods 
have been developed that are applicable to a wide class of observables.

At lower energies, BEVALAC, SIS, AGS, and SPS, the physics of nuclear
collisions is now known to be dominated by the non-equilibrium dynamics of the
confined {\em intermediate} phase of QCD, known as
hadronic resonance matter\cite{Bass:1998vz}.  Unfortunately, even lattice QCD methods
are not yet powerful enough to predict the dynamics or thermodynamics
of this intermediate form of matter.  For low temperature nuclear matter,
effective quantum hadro-dynamic theories (QHD, Chiral Perturbation
Theory) have successfully been constructed incorporating the known
(Lorentz and Chiral) symmetries of QCD. However, at moderate
temperatures where matter is in the hadron resonance excitation region
(below the deconfinement temperature) no quantitative theory yet exists.
This handicaps the interpretation of data at such lower
energies in terms of fundamental QCD properties. Many phenomenological
hadronic dynamical models have been advanced to help interpret data at lower energies, 
but as yet they have not evolved into a consistent effective theory.  
The existing huge data base from lower energies provides valuable
information for further development of such an effective
theory. However, RHIC energies provide for the first time the possibility of
exploring a qualitatively new kinematic regime where the uncertainties due to our as yet
incomplete understanding of the intermediate hadronic resonance phase of QCD may
be minimized.

The case that we present in this report, based on the published data
from RHIC, about the Quark Gluon Plasma and the Color Glass Condensate
is predicated by the firm root of these concepts in first principles
in QCD.  As such, these concepts must be tested thoroughly through a
wide array of observables to see whether they are consistent with
available measurements, and that they are robust in their predictions.
It was a priori not at all obvious whether any observables exist at
RHIC that can be described quantitatively by the QGP or CGC concepts.
This is where a healthy bit of experimental ``luck'' was essential in
order to find  the  ``needles in the haystack'' that  are least
distorted by uncertain non-equilibrium hadronic final state
dynamics.  We make the case in this report that a few sharp needles have been
found through three convergent lines of empirical evidence that point
to the discovery of a new strongly coupled QGP and its
source, the saturated CGC.

It is also important to understand when an approximation to the QCD
theory breaks down.  Only special limits of QCD can be quantitatively
predicted at this time: (1) long wavelength QCD thermodynamics, (2) very short wavelength
pQCD, and (3) very high energy CGC coherence. New quark coalescence techniques
 are being developed to address the intermediate wavelength observables.  
Any breakdown of a given approximation 
must be understood for solid reasons.  
One case in point is applying the QGP hypothesis to
observables (such as radial flow) which are strongly influenced by the
uncertain late time dynamics of hadronic matter in heavy ion collisions.
Another case is at the earliest formation times, when the description
changes between the initial coherent CGC and the produced thermal QGP.
In the latter case, one needs to determine directly from careful
control experiments, which approximation to QCD is better suited to
the phenomena under study. 
In this report we will discuss both
the strengths and limitations of the present theoretical understanding
of the RHIC discoveries.

\section{What is the Quark Gluon Plasma?}

Quantum Chromodynamics (QCD) 
predicts the existence of a {deconfined } form of matter called a Quark Gluon
Plasma (QGP), in which the quark and gluon degrees of freedom normally
confined within hadrons are mostly liberated\cite{Collins:1974ky}.  
This transition occurs
when the energy density of matter is of the order of that of matter
inside a proton.  This density is about an order of magnitude larger
than the energy density inside of atomic nuclei, that is about
$1-10~GeV/fm^3$.

To get a more accurate determination of the energy of this transition
to a Quark Gluon Plasma, QCD can be numerically studied using numerical
methods\cite{Allton:2003vx}-\cite{Fodor:2004nz}.  Such computations show that
there is a rapid rise of the energy density, $\epsilon (T)$, of matter
when the temperature reaches $T\approx T_c\sim 160$ MeV.  The energy
density changes by about an order of magnitude in a narrow range of
temperatures $\Delta T \sim 10-20~ MeV$ as can be seen from Fig. 1.

\begin{figure}[htb]
        
   \centering
       \mbox{{\epsfig{figure=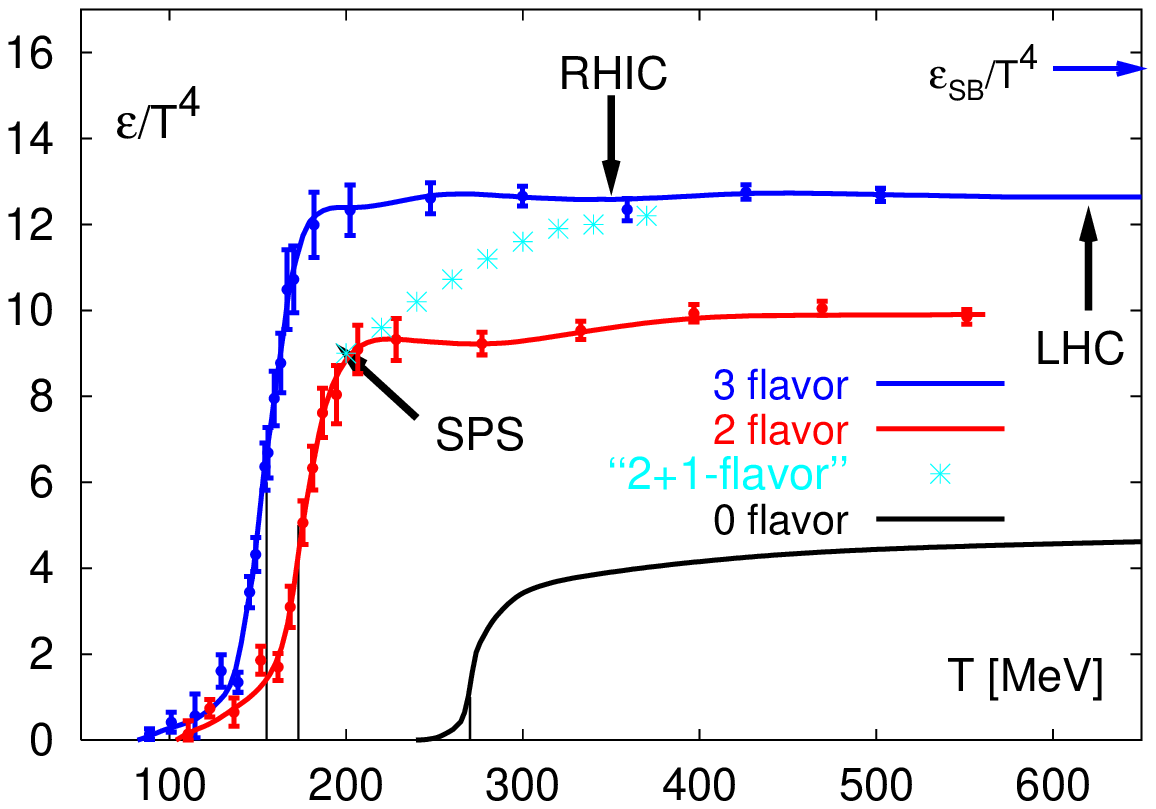,
        width=0.450\textwidth}}\quad
             {\epsfig{figure=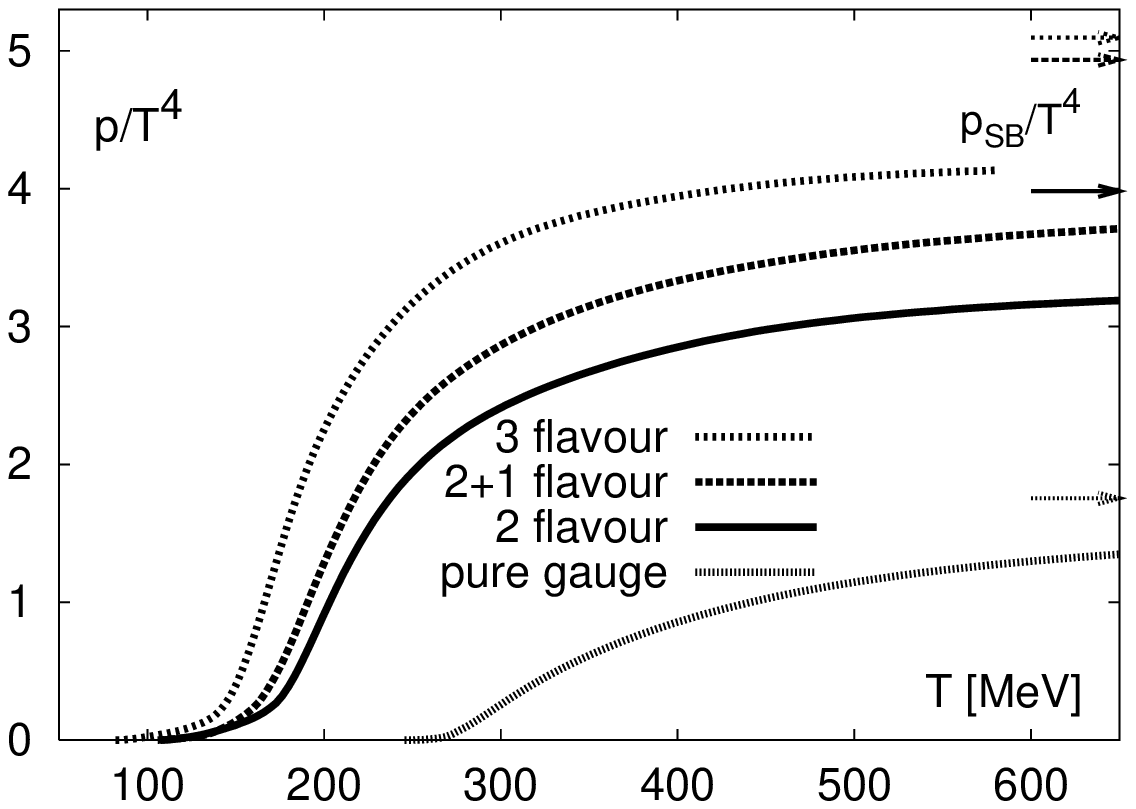,
        width=0.450\textwidth}}}
\begin{minipage}[hbt]{6in}
\caption{
{\small (a) The energy density as a function of temperature
scaled by $T^4$ from lattice QCD \protect{\cite{Allton:2003vx}}. 
Various number of species of quarks are considered.
The realistic case is for $2+1$ flavors.  An estimate of the typical 
temperature reached at SPS and RHIC, and 
estimated for LHC is included in the figure.(b)  The pressure as a function
of temperature scaled by $T^4$.  Note that the pressure is continuous in 
the region where there is a sharp change in the energy density.}
\label{eos} }
\end{minipage}

\end{figure}

One can understand this transition as a change in the 
number of degrees of freedom of the system.  Far below $T_c$, the active  hadronic degrees
of freedom are limited to three, corresponding to a dilute gas of 3 charge states of pions.
Above $T_c$, 8 colors of gluons times two helicity degrees of freedom become activated. 
In addition, there
are $N_f(T)\approx 2-3$ active light flavors of quarks 
(for T not far above $T_c$).
Each flavor has equal number of quarks and 
antiquarks when the chemical potentials
vanish, and each have two spin states and three colors. Therefore, 
there are $g_{q\bar{q}}(T)\approx 24-36$
quark degrees of 
freedom in a QGP.  In the quark gluon plasma phase,  
there are then about 40 - 50 internal
degrees
of freedom in the temperature range $(1-3)T_c$,
while the low temperature and vanishing chemical potentials
the pion gas has 3.  Since the energy
density, pressure and entropy are all roughly proportional to the number of 
degrees of freedom, one understands this rapid change in the energy density
over a narrow range of temperature as a change in the degrees of freedom 
between the confined and deconfined worlds. 

The system above $T_c$ is called a plasma because the degrees of freedom carry 
the non-Abelian analog of charge as in ordinary plasmas. Just as there are different
regimes of ordinary plasmas, the QGP plasma also has weakly coupled and strongly coupled
limits. At extremely high temperatures, the asymptotic freedom property of QCD 
predicts that it will be weakly coupled\cite{Collins:1974ky}. 
For moderate temperatures $(1-3)T_c$ accessible at RHIC, on the other hand,
the plasma is predicted by nonperturbative lattice techniques to be strongly coupled
- even though the thermodynamic variables are near the ideal Stefan Boltzmann limit.
{Thus, nonperturbative correlations beyond dielectric phenomena persist
in the QGP well beyond $T_c$.}

The transition between the confined hadronic world and the deconfined QGP world
may or may not be a phase transition in the strict statistical mechanical
sense. Stricly speaking, a phase transition requires
a mathematical discontinuity in the energy
density or one of its derivatives in the infinite volume limit.  
The QGP transition may in fact be 
a ``crossover'', or rapid change, as is suggested numerical computation 
and a number of theoretical arguments.  Nevertheless, the change
as measured in numerical computation is very abrupt as seen in
Fig. \ref{eos}. 

We shall refer to matter in the transition region as a ``mixed
phase''.  If there were a strict statistical mechanical first
order phase transition, then matter in this region would be a
mixture of hadronic gas phase and quark gluon plasma domains, as is the case
when there is phase coexistence between water and ice.  If there is a
rapid crossover, then many of the bulk properties of the system, which are
determined by the relation between energy density and pressure, are to
a good approximation similar to those when there is a strict phase
transition.

\begin{figure}[h]
\centering
\mbox{{\epsfig{figure=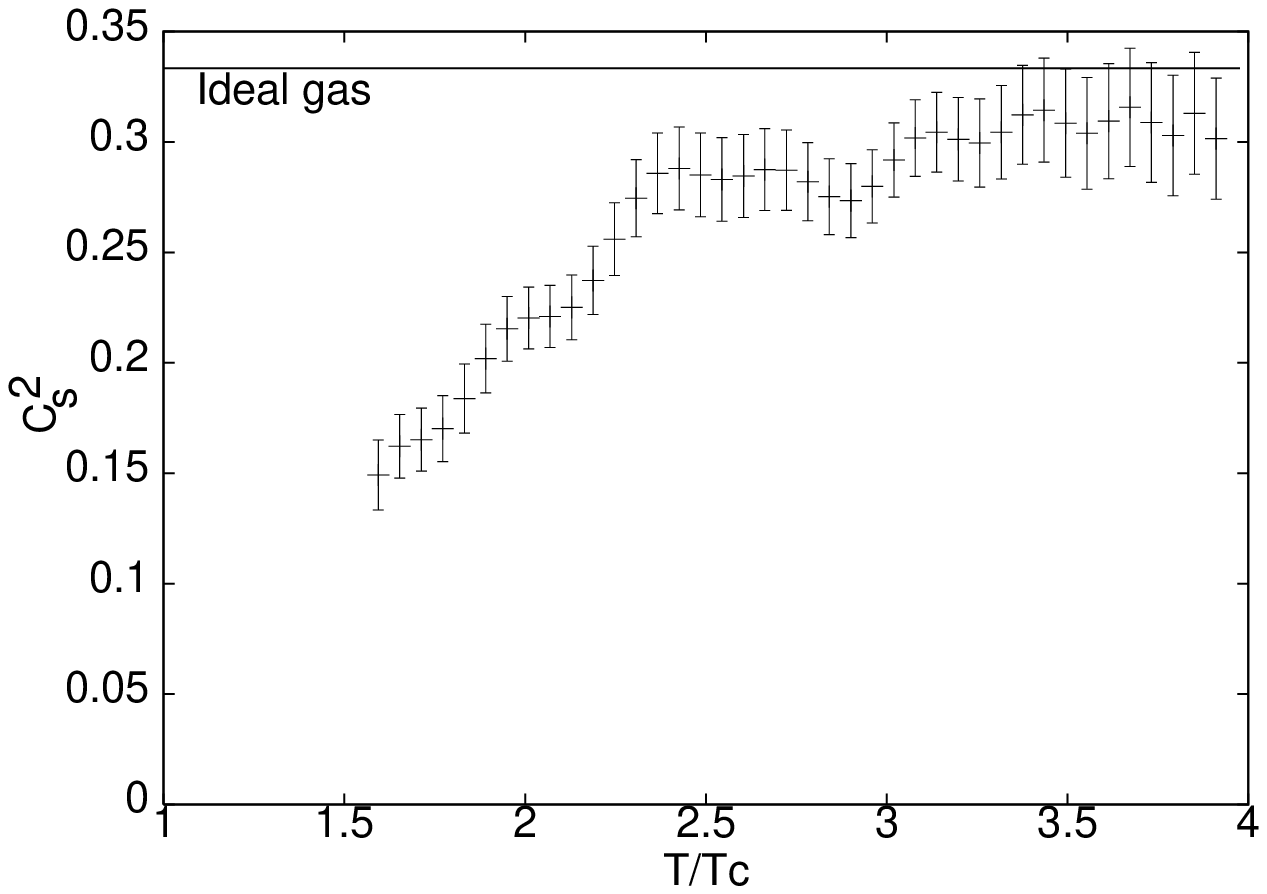,
        width=0.40\textwidth}}\quad \vspace{1in}
             {\epsfig{figure=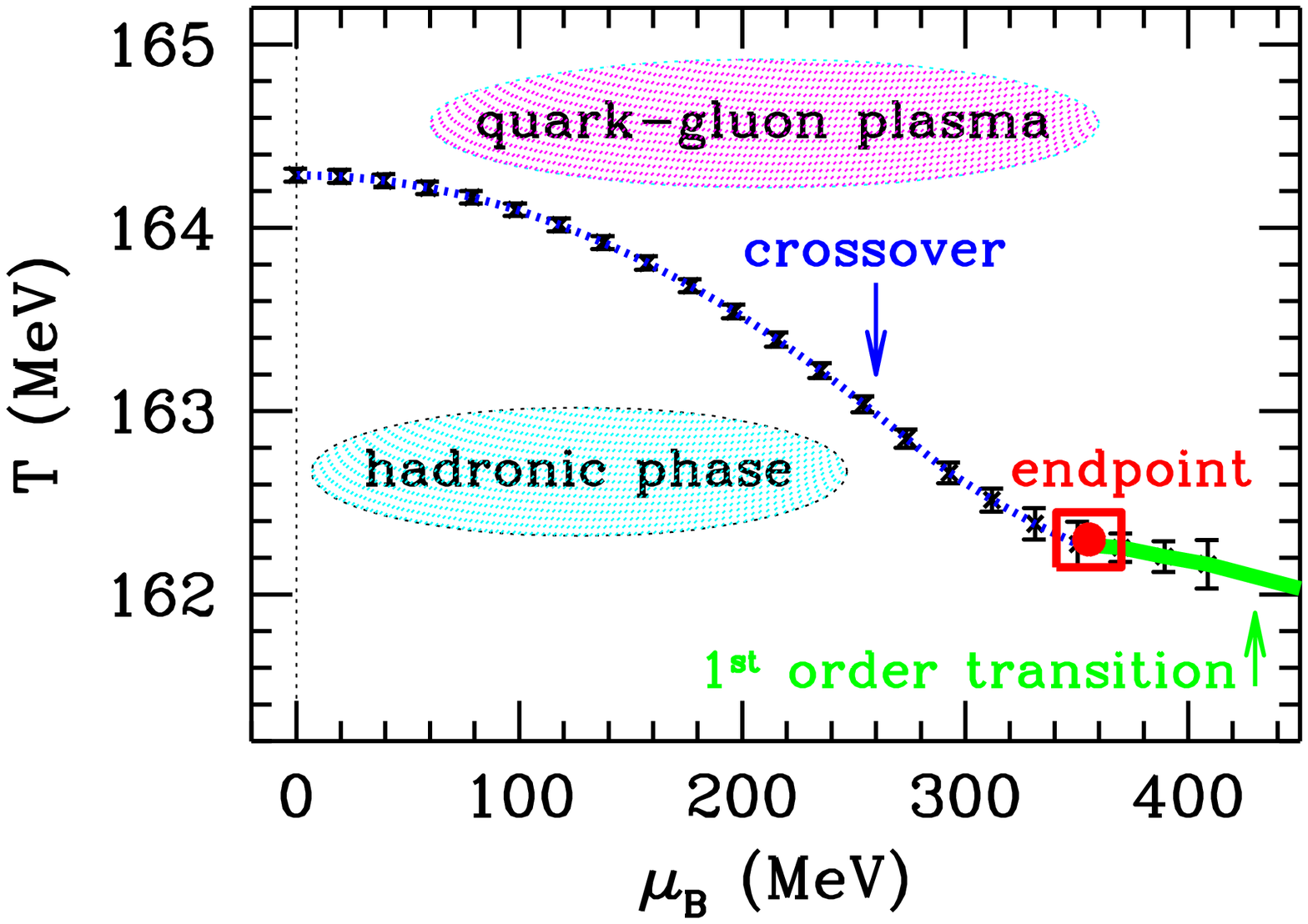,
        width=0.450\textwidth}}}
\begin{minipage}[hbt]{6in}
\caption{ {\small Important features of the QGP equation of state. (a)
The speed of sound\protect{\cite{Bernard:1996cs,Gupta:2003be}} 
$c_s^2=d\epsilon/dP$ drops below $1/3$ for $T<2T_c \approx300$ MeV.
(b) Right panel shows a current estimate of the location of the 
tri-critical point at finite baryon density\protect{\cite{Fodor:2004nz}} 
}\label{qgpfig2}
}
\end{minipage}

\end{figure}

In the transition region, the energy density changes by roughly an
order of magnitude, but the pressure is continuous and varies slowly. 
 The sound
velocity, $c_S^2 = dP/d\epsilon$ must therefore become very small in
this range of energy densities\cite{Bernard:1996cs,Gupta:2003be}.  It
is very difficult to generate pressure gradients and do mechanical
work in the mixed phase region, since as we vary the energy density we
generate little change in pressure.  We will call a relation between
energy density and pressure (an equation of state) stiff when the
sound velocity is big ($c_s$ of order the speed of light) and soft
when it is small.  The quark gluon plasma equation of state is stiff
at high temperature, but becomes soft near $T_c$.  The sound velocity
as a function of temperature as determined by numerical computation is
shown in Fig. \ref{qgpfig2}a.  It is expected that hadronic
matter again becomes stiffer below $T_c$, and eventually
softens as the temperature tends  to zero.
Unfortunately, lattice QCD is still not powerful enough to predict accurately the
thermodynamic properties of hadronic matter below $T_c$. Preliminary 
results\cite{Karsch:2003zq} (still with rather large pion mass $\sim 700$ MeV)
 are roughly consistent with a heavy hadronic resonance gas
equation of state, but the thermodynamics
of the confined phase of QCD remains an open problem. This fact again underscores
the necessity of concentrating on those (few) observables that are least influenced by the
poorly known hadronic phase dynamics.

The relationship between the energy density and pressure determines
the evolution of matter from 
a given  initial condition {\em IF} local equilibrium is maintained.
This then can determine  experimentally by measuring
``barometric'' observables. 
The softening of the QGP equation 
of state near $T_c$
is the key feature that can be looked for
in  the collective hydrodynamic
flow patterns produced when the plasma expands.

Another distinctive feature of the QGP phase diagram is shown in the 
right panel in 
Fig.(\ref{qgpfig2}). Recent numerical calculations\cite{Fodor:2004nz}
have begun to show evidence that the QGP may have a phase transition
in the strict statistical mechanical sense, if one also allows the system
to have a high baryon number density as well as high temperature.
\cite{Halasz:1998qr,Rischke:2003mt} 
In the plot on the right hand side of Fig. \ref{qgpfig2},
the possible phases of QCD are shown as  a function of both temperature
and a measure of the baryon number density, $\mu_B$.  At low baryon
number density, there is probably a rapid crossover from the
low density hadronic world to that of the quark gluon plasma.
At higher baryon number density, there is quite likely a first order phase 
transition. The computations on this issue are still work in progress  and
extrapolation to realistic quark masses are very difficult, so the position
of the endpoint of first order phase transitions 
is still rather uncertain. There  is an ongoing program at the SPS
concentrating on lower energies 
to search for possible signatures of high baryon density quark plasma transition. 

The quark gluon plasma has the property that quarks and gluons are no
longer confined inside of hadrons.  This is unlike zero temperature and baryon
number density QCD.  The transition between the confined and deconfined world 
may provide us insights into the nature of confinement.

In the QCD equations of motion the small mass of the ``light'' u,d quarks 
can be neglected and the system possesses a special chiral symmetry. 
This chiral symmetry prevails at high temperature, in spite of the fact
that quarks acquire a collective dynamical
mass $\sim gT$ as do gluons. Below $T_c$,  not only do quarks and gluons become confined
into hadrons, but the chiral symmetry of QCD is also broken as manifested by the
appearance of the 3 light pion degrees of freedom while the quarks are
bound in heavy nucleons.

\vspace{2ex}
The Quark Gluon Plasma is important to search for at RHIC because:

\begin{itemize}   

\item{It is the ultimate, primordial form of QCD matter at high temperature or 
baryon number density (at least up to the electro weak scale at about $T_{EW}
\sim 10^3 T_c$).}

\item{It was present during the first few microseconds of the Big Bang according to current cosmology.}

\item{It may occur naturally in supernovae, gamma ray bursts and neutron stars
as matter at high baryon number density and relatively low temperature.}

\item{It provides an example of phase transitions which {may} occur at a variety 
of higher temperature scales in the early universe.}

\item{It may provide us important information concerning the
origin of mass for matter, and how quarks are confined into hadrons.}

\end{itemize}

\section{What is the Color Glass Condensate?}

The ideas for the Color Glass Condensate are motivated by HERA data 
on the gluon distribution function shown in 
\begin{figure}[htb]
    \centering
       \mbox{{\epsfig{figure=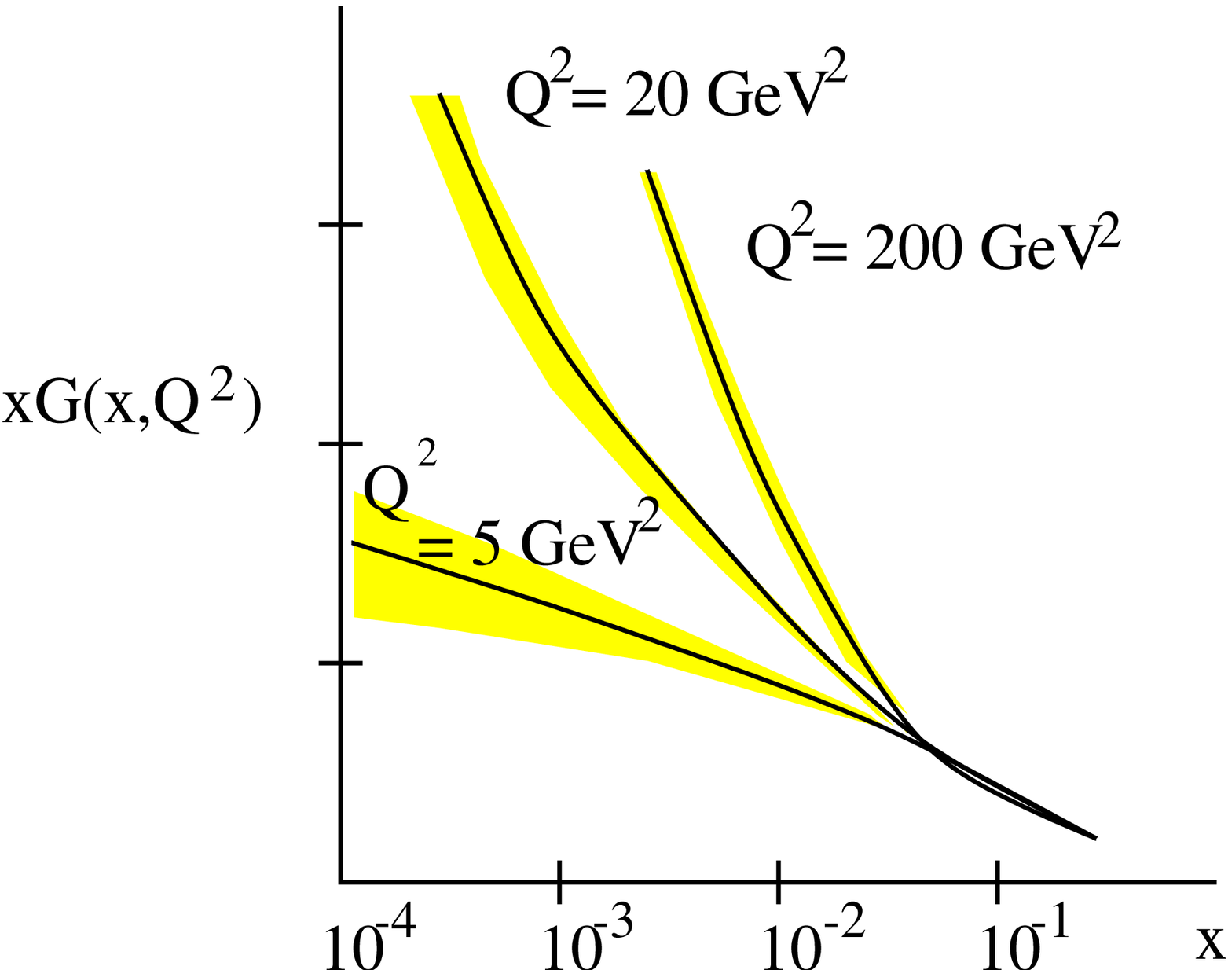,
        width=0.45\textwidth}}\quad
             {\epsfig{figure=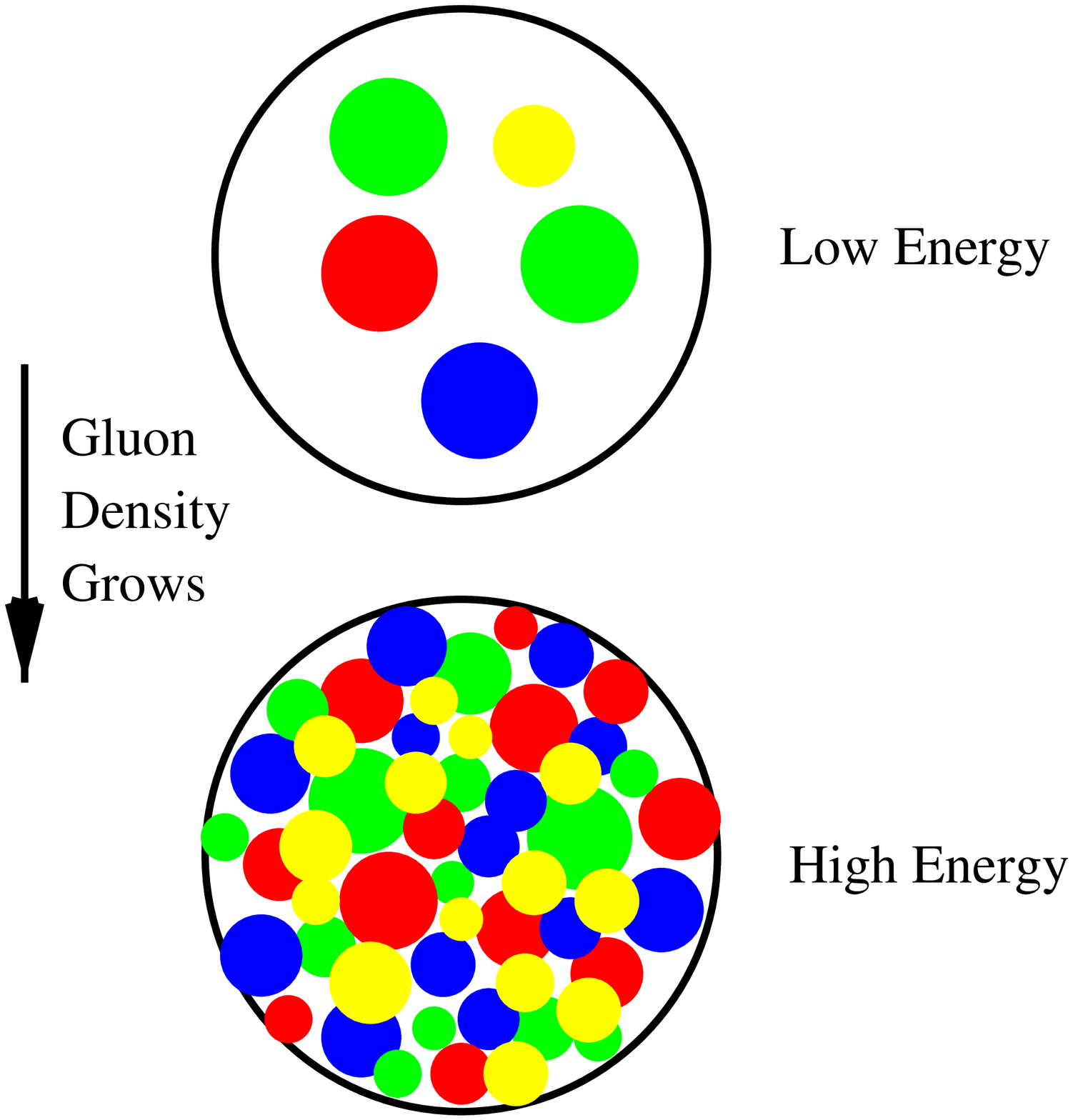,
        width=0.45\textwidth}}}
\begin{minipage}[hbt]{6in}
        \caption{{\small (a)The HERA data for the gluon distribution function
as a function of x for various values of $Q^2$. (b) 
A physical picture of the low x gluon density inside a hadron
as a function of energy } 
\label{heradata}}
\end{minipage}
\end{figure}
Fig. \ref{heradata}(a)
\cite{hera}. 
The gluon density , $xG(x,Q^2)$, rises rapidly 
as a function of decreasing fractional momentum, x, or increasing
resolution, $Q$.  {This rise in the gluon density ultimately owes
its origin to the non-Abelian nature of QCD and that the gluons
carry color charge.} At higher and higher energies, smaller $x$ and larger $Q$
become kinematically accessible.   
The rapid rise with $\log (1/x)$ was expected in a variety of 
theoretical works\cite{glr}-\cite{bfkl}. Due to the intrinsic non-linearity of QCD,
gluon showers generate more gluon showers -  producing an exponential avalanche
toward small $x$.
The physical consequence of this exponential growth is that the
density of gluons per unit area per unit rapidity 
of any hadron including nuclei 
must increase rapidly as $x$ decreases 
\cite{mv}.  This follows because the transverse size, as seen via the total cross 
sections,  rise more  slowly at high energies than 
the number of gluons. This is illustrated  in Fig. \ref{heradata}(b). 
The non-linearity of gluon interactions, however,  led to
the conjecture that the transverse density of gluons 
{measured at some fixed $Q^2$ resolution scale} should 
{eventually} become limited ,
that is, there is gluon saturation at sufficiently high energies. 
\cite{glr}-\cite{mq}, \cite{mv}

The low x gluons  are closely packed together in the transverse direction, therefore,
as in the thermal case at extreme temperatures or chemical potentials,  the 
strong interaction 
strength must become weak, $\alpha_S \ll 1$. 
Weakly coupled systems 
should be possible to understand from first principles in $QCD$ \cite{mv}.  

This dense but weakly coupled system is called a Color Glass Condensate for reasons 
we now enumerate:\cite{ilm}
\begin{itemize}
\item {\bf Color}  The gluons which make up this matter are colored.
\item{\bf Glass} The gluons at small x are generated from gluons at larger
values of x.  In the infinite momentum frame, these larger momentum gluons
travel very fast and their natural time scales are Lorentz time dilated.  
This time dilated scale is transferred to the low x degrees of freedom
which therefore evolve very slowly compared to natural time scales.  This
is the property of a glass.
\item{\bf Condensate} The dimensionless transverse phase space density 
\be
\rho = {1 \over {\pi R^2}}{{dN} \over {dyd^2p_T}}
\ee
saturates at a value $\rho = c/\alpha_S(Q_s)$, where $c$ is a constant of order unity. 
Saturation is due to the competition between
extra gluon radiation originating from the source current, $\propto \rho$, and
non-linear gluon fusion, $\propto -\alpha_S \rho^2$, that reduces the number of gluons
at high density. 

Because $\alpha_S << 1$,
this means that the quantum mechanical states of the 
system associated with the condensate are multiply occupied.  They are 
highly coherent, and share some properties of Bose condensates.
The gluon occupation factor is very high, of order $1/\alpha_S$, but it is
only slowly (logarithmically) increasing when further increasing the
energy, or decreasing the transverse momentum. This provides saturation
and cures the infrared problem of the traditional BFKL approach \cite{im2001}.

\end{itemize}

There is a critical momentum scale, $Q_s(x,A)$, which controls the occupation number
through $1/\alpha_S(Q_s)$. This is called 
the  CGC saturation scale. The transverse phase space density, $\rho$,
is constant only up to $p_T < Q_s(x,A)$. For wavelengths, $1/p_T$, much smaller
than $1/Q_s$, the coupling becomes even weaker and perturbative QCD 
predicts that $\rho\propto
\alpha(p_T)/(R^2 p_T^2)$. 

Note that $Q_s$ plays a role in the CGC form of matter 
similar to what $T_c$ plays in the QGP form of matter. Both
delineate two different phases of matter. However, in the QGP case,
the critical scale, $T_c\sim \Lambda_{QCD}$, is small and the matter on both sides
of the transition remain in the
strong coupling sector of QCD! In the
CGC case, on the other hand, it is possible to find kinematic regimes of $x$ 
that depend on
$A$, where the CGC can be weakly coupled but in nonlinear regime due to high occupation
numbers. 

The crux of the search for the CGC therefore is to locate those kinematic regimes
where the corrections to the weak coupling methods can be controlled. 
The  non-linear dependence of $Q_s$ on $x$ and $A$ 
will be discussed further in section 6.  {We note here only that both
as $x$ becomes small and $A$ becomes large, $Q_s$ grows.  The saturation
momentum itself does not saturate.  The gluon distribution function for
resolution scale $Q \le Q_s$ grows slowly and saturates, while the gluon distribution
function for $Q \ge Q_s$ grows rapidly.  The physics is easy to understand:
As more gluons are added to the hadron, they have to go to the unsaturated
region since the saturated region is already densely packed.}

The Color Glass Condensate is important to search for at RHIC because:
\begin{itemize}   

\item{It represents the  universal form
of high energy QCD wavefunctions at small $x$,  not only of hadrons
but heavy nuclei as well. }

\item{It is a new form  of matter because the gluons inside the
hadron are separated by distance scales small compared to the hadron size, they
evolve on time scales long compared to microphysics time scales, and the CGC itself
is specified by a special new density matrix that encodes the non-linear
virtual fields of QCD in the high energy limit. }

\item{It {already begins to become important at $x \sim 10^{-2}$}
  in protons at HERA, 
and and should begin to become important at
comparable values of $x$  at RHIC.}

\item{It provides a rigorous QCD theoretical description of the initial state 
in $A+A$ from which the QGP must evolve.}

\item{It can be tested directly at RHIC via $p+A$ or $D+A$, where
no final state interactions and hence no QGP formation can occur. }

\end{itemize}

\section{The Space-Time Picture of Ultra-Relativistic
Nuclear Collisions}

\begin{figure}[ht]
    \begin{center}
\includegraphics[width=.50\textwidth]{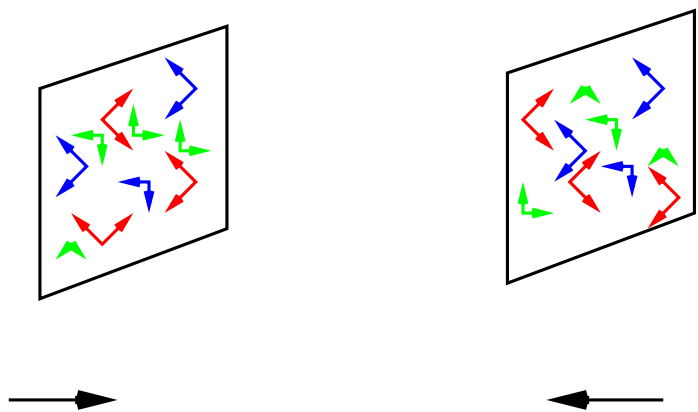}
\begin{minipage}[hbt]{6in}
        \caption{{\small The collision of two sheets of colored 
glass.  The arrows represent the polarization vectors for the
gluons which live on the sheets, and their colors correspond to the
different colors of gluons}\label{sheetonsheet}}
\end{minipage}
    \end{center}
\end{figure}

Heavy ion collisions at ultrarelativistic energies are visualized in Fig. 
\ref{sheetonsheet} as the collision of two sheets of colored glass.\cite{aa}
The nuclei appear as sheets at ultrarelativistic energies because
of Lorentz contraction. The CGC gluons are shown as
vectors which represent the polarization of the gluons, and by colors 
corresponding to the various colors of gluons.

At ultrarelativistic energies, these sheets pass through one another.
In their wake is left melting colored glass, which eventually materializes as 
quarks and gluons.  These quarks and gluons would naturally form in their
rest frame on some natural microphysics time scale.
For the saturated color glass,  this proper formation
time scale, $\tau_0$, is of 
order the inverse saturation momentum. At RHIC,  $1/Q_s\sim 0.2$  fm/c 
$\sim 5\times 10^{-25}$ sec. Note that $0.2\;{\rm fm/c}$ is also comparable
to the natural crossing time
of two 10 fm nuclei, each contracted by a gamma factor 100, in the center of mass frame
 at RHIC. 
For particles with 
a large momentum or rapidity along the beam axis, this time scale is Lorentz dilated.
This means that the slow (smaller rapidity) 
particles are produced first towards the center of
the collision regions and the fast (larger rapidity) 
particles are produced later further away
from the collision region.  

\begin{figure}[ht]
    \begin{center}
  \includegraphics[width=.40\textwidth]{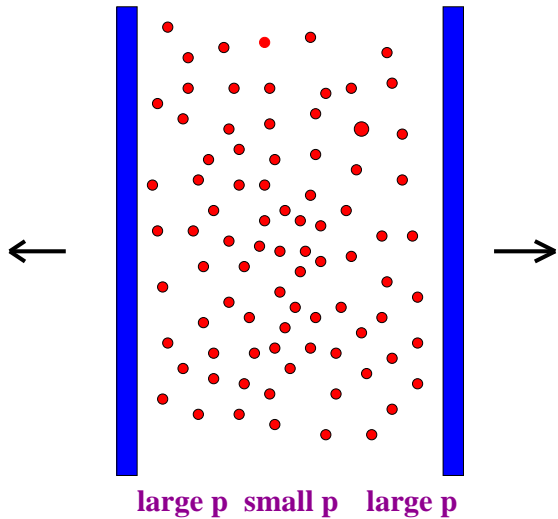}
        \caption{{\small Particles being produced after the 
collision of two nuclei.}\label{collision}}
    \end{center}
\end{figure}

This Bjorken ``inside-out''  correlation\cite{Bjorken:uj}
 between space and momentum is similar to what happens
to matter in Hubble expansion in cosmology.  The stars which are further away
have larger outward velocities.  This means that the matter produced at RHIC, 
like the universe
in cosmology is born expanding. One important difference is that the
``mini-bang'' at RHIC is born with one dimensional Hubble flow along the collision axis,
while the Big-Bang is three dimensional.  This is shown in Fig. \ref{collision}

As this system expands, it cools. On some time 
scale $\tau_{eq}> \tau_0 $ the produced quarks
and gluons may thermalize. It is by no means obvious
that local equilibrium can be reached on the short time scale $\sim (1-2) R/c \sim 10$ fm/c
available in nuclear collisions. {\em IF} local equilibrium is reached early
with $\tau_{eq}< 1$ fm/c , then
the QGP can develop collective flow according to the laws of hydrodynamics.
In any case, when the energy density decreases due to expansion
below about $1$ GeV/fm$^3$, the QGP must begin to hadronize
through some mixture of hadrons and quarks and gluons.  Eventually, 
 all the quarks and gluons must become confined
into hadrons that may still interact as hadronic matter before
being detected. 

\begin{figure}[h]
    \begin{center}
 \includegraphics[width=0.60\textwidth]{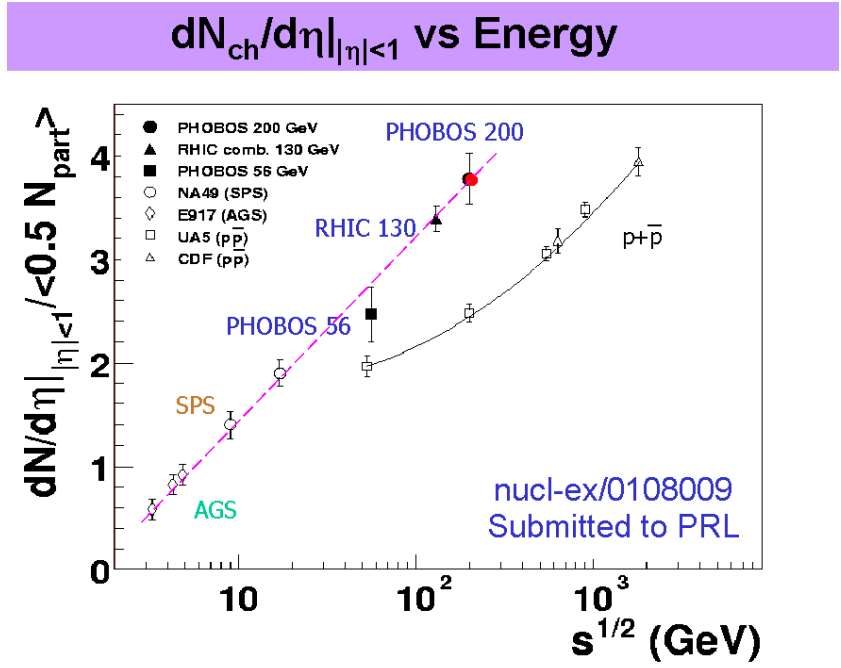}
\begin{minipage}[hbt]{6in}
        \caption{{\small The particle multiplicity as a function of c.m. energy, $\surd s$,
per nucleon pair at RHIC\protect{\cite{Back:2001ae}} and lower AGS and SPS energies.  
                            }\label{dndye}   }
\end{minipage}
 \end{center}
\end{figure}

The  particle multiplicity as a function of energy has been measured at 
RHIC\protect{\cite{Back:2001ae}}, as shown in Fig. \ref{dndye}. 
Combining the multiplicity data together with the measurements of transverse
energy or of typical particle transverse momenta, one can determine the
energy density of the matter when it decouples.\cite{dndyphobos}  
One can then extrapolate
backwards in time using 1 dimensional 
(Bjorken) expansion, since decoupling occurs soon after 
the matter  begins
to expands three dimensionally.  We can extrapolate backwards only until $\tau_0$, 
when the matter is formed from the ``shattered''  Color Glass.

To do this extrapolation we use that the proper density
of particles falls as $N/V \sim 1/\tau$ during 1 dimensional expansion.  
If the particles expand without
interaction and work, then the energy per particle remains
 constant ( and $E/V(\tau_0)\approx (dE_T/dy)/(\tau_0 \pi R^2)$ 
in terms of the final observed transverse energy per unit rapidity).  If the particles
thermalize, then $E/N(\tau) \sim 3T(\tau)$, and 
the {\em entropy} rather than the energy per particle
remains constant.  For a massless
gas, the temperature then falls  as $T \sim \tau^{-1/3}$.  
For a gas which is not massless or not in perfect equilibrium, 
the temperature falls somewhere in the range $T_o
> T(\tau) > T_o (\tau_{eq}/\tau)^{1/3}$ This 1 dimensional expansion continues
until the system begins to feel the effects of finite size in the 
transverse direction, and then rapidly cools through three dimensional
expansion.  
\begin{figure}[h]
    \begin{center}
   \includegraphics[width=0.80\textwidth]{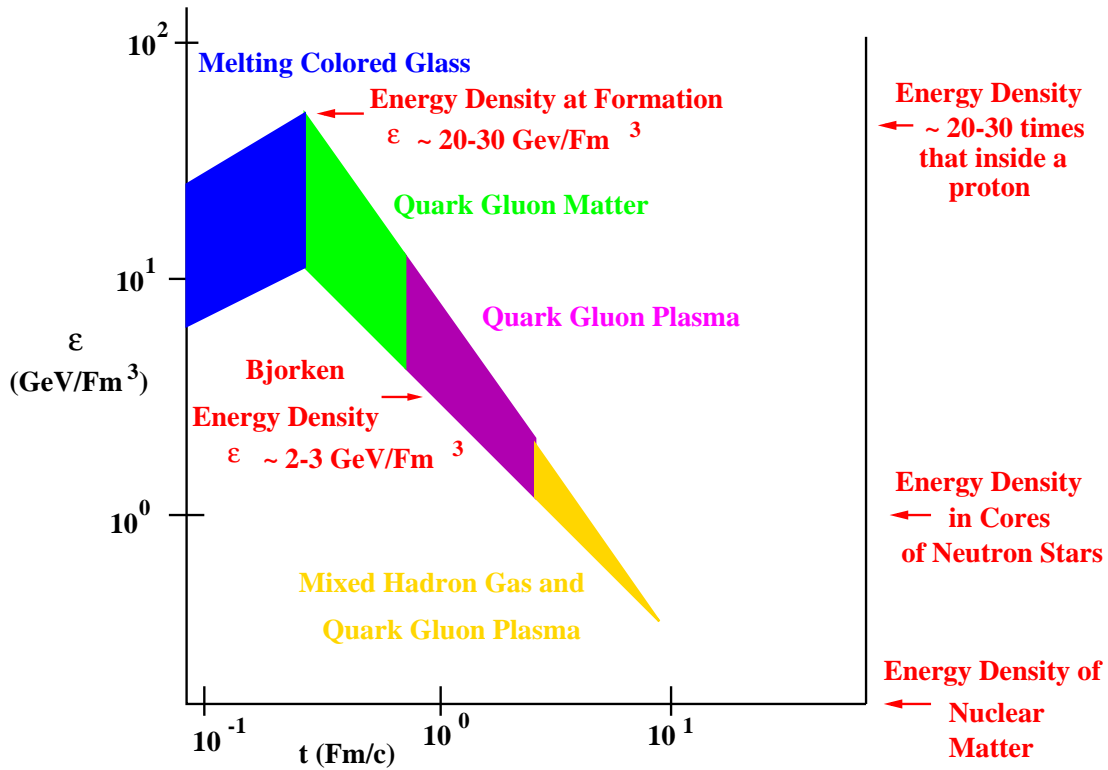}
        \caption{Bounds on the energy density as a function of time
in heavy ion collisions.  
                            }\label{times}   
 \end{center}
\end{figure}
We shall take a conservative 
overestimate of this time to be of order $t_{0} \sim 0.3$ fm/c  
The
extrapolation of the energy density backwards is bounded by $\epsilon_{f} (t_f/t)
< \epsilon(t) < \epsilon_f (t_f/t)^{4/3}$.  The lower bound is
that assuming that the particles do not thermalize and their typical
energy is frozen.  The upper bound assumes that the system thermalizes
as an ideal massless gas.  
These bounds on the energy density is shown in Fig. \ref{times}.
On the left axis is the energy density and on the bottom axis is time.
The system begins as a  coherent Color Glass Condensate, then melts to Quark
Gluon Matter which may eventually thermalize to a Quark Gluon Plasma.
At a time $\sim 3 fm/c$, the plasma becomes a mixture of quarks, gluons 
and 
hadrons which further expand together.  

At a time of about $10 ~fm/c$, the
system falls apart and decouples.  At a time of $t \sim 1~fm/c$,
the estimate we make is identical to the Bjorken energy density estimate,
and this provides a lower bound on the energy density achieved 
in the collision.  (All estimates agree that by a time of order $1 ~fm/c$,
matter has been formed.)  The upper bound corresponds to assuming
that the system expands as a massless thermal gas from a melting
time of $.3~fm/c$.  (If the time was reduced, the upper bound would  
be increased yet further.)
The bounds on the  initial  energy density 
are therefore
\be
        2-3 ~GeV/fm^3 \lton \epsilon \lton 20-30 ~GeV/fm^3
\ee
where we included a greater range of uncertainty in the upper limit
because of the uncertainty associated with the formation time.
The energy density of nuclear matter is about $0.15~GeV/fm^3$, and
even the lowest energy densities in these collisions is in excess of this.
At late times, the energy density is about that of the cores of neutron stars,
$\epsilon \sim 1 ~GeV/fm^3$.  

\vspace{0.2in}

{\bf We conclude that based on the observed multiplicity at RHIC
alone, the initial energy densities achieved in RHIC collisions 
can be high enough to produce a quark gluon plasma.}

\section{Empirical Evidence for QGP at RHIC}

In this section, we discuss RHIC data which show that
the matter produced in central collisions in RHIC becomes well thermalized,
and behaving in a way consistent with theoretical expectations from QCD.

\subsection{Collective Flow}

The identification of a new form of ``bulk matter''  requires the 
observation of novel and uniquely different
collective properties from ones seen before. In heavy ion reactions
the flow pattern of thousands of produced hadrons is the primary
observable used to look for novel collective phenomena\cite{Hofmann:by}, \cite{Stocker:bi}-
\cite{Reisdorf:1997fx}. 
The collective flow properties
 test two of the conditions necessary
for the validity of the QGP hypothesis.

The first is the degree of thermalization.  Nothing is yet known from lattice QCD
about far off equilibrium dynamics of a QGP. However, the
evolution of matter from some initial condition can be computed via the equations
of viscous relativistic hydrodynamics if local equilibrium is maintained. 
 These equations can be further approximated   by
perfect (Euler) fluid equations when the corrections due to viscosity can be neglected.
Such viscous corrections can be neglected  when scattering mean free paths are small
compared to the scale of spatial gradients of the fluid.

The second condition is the validity of the numerically determined equation of state
or relationship between energy density and pressure.
The required input for perfect fluid
hydrodynamical equations is the equation of state. With a specific
initial boundary condition, the future evolution of the matter can be then
predicted. We shall show that the data on elliptic
flow confirms the idea that to a very good approximation, 
local thermal equilibrium is reached at RHIC energy and that the flow
pattern is entirely consistent with
numerical determinations of the equation of state from QCD.

The different types of collective 
flows are conveniently quantified in terms of the first 
few Fourier components of the azimuthal angle (angle around the beam axis 
for the collision) distribution.~\cite{Voloshin:1994mz,Ollitrault:bk}, $v_n(y,p_T,N_p,h)$,
of the centrality selected triple differential inclusive distribution
of hadrons, $h$. The centrality or impact parameter range is usually
specified by a range of associated multiplicities, from which
the average number of participating nucleons, $N_p$, can be deduced.
The azimuthal angle of the hadrons are measured relative to 
a globally determined estimate for the collision reaction plane angle 
$\Phi(M)$. The ``directed'' $v_1$ and ``elliptic'' $v_2$ flow components
\cite{Reisdorf:1997fx}-\cite{Ollitrault:bk},
\cite{Voloshin:1999gs}-\cite{Adler:2003kt}
are
readily identified from azimuthal dependence
\begin{eqnarray}
\frac{dN_h(N_p)}{dydp_T^2d\phi}
=\frac{dN_h(N_p)}{dydp_T^2} \frac{1}{2\pi}(1 &+& 2 v_1(y,p_T,N_p,h) 
\cos\phi  \nonumber
\\   &+&  2 v_2(y,p_T,N_p,h) \cos 2 \phi + \cdots ) \;\;.
\label{floweq}
\end{eqnarray}
The first term in the above equation also contains information
about flow.  Produced particles should have their momentum spectrum 
broadened in heavy ion collisions relative to the case for 
proton-proton collisions.  Because flow is due to a collective velocity,
the flow effects should be largest for the most massive particles,
and therefore the  mass dependence is a powerful diagnostic tool.  

\begin{figure}[h]
\centering
\includegraphics[height=0.35\textheight,width=0.45\textwidth,clip]
{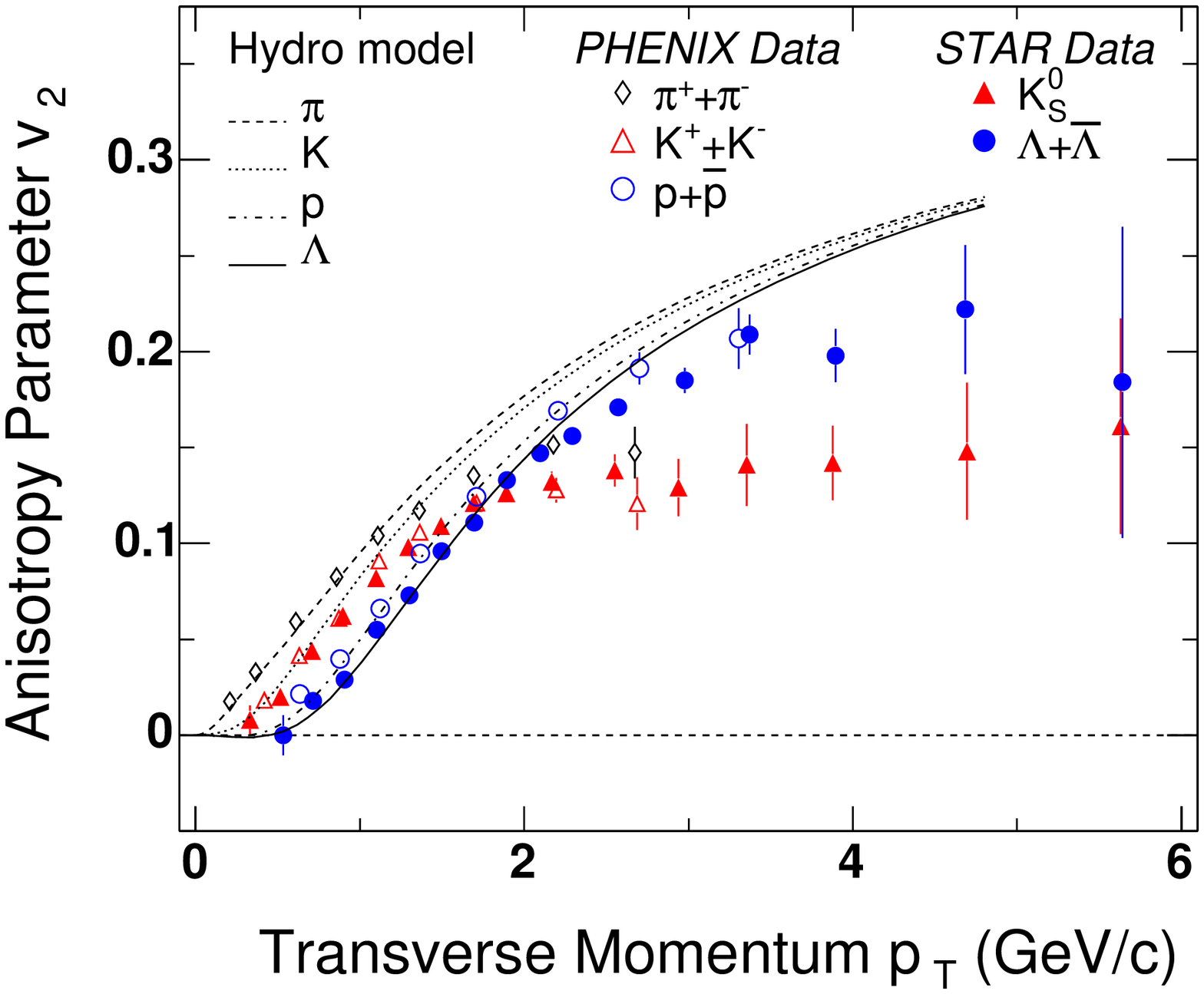}
\hfill
\includegraphics[height=0.35\textheight,width=0.54\textwidth,clip]
{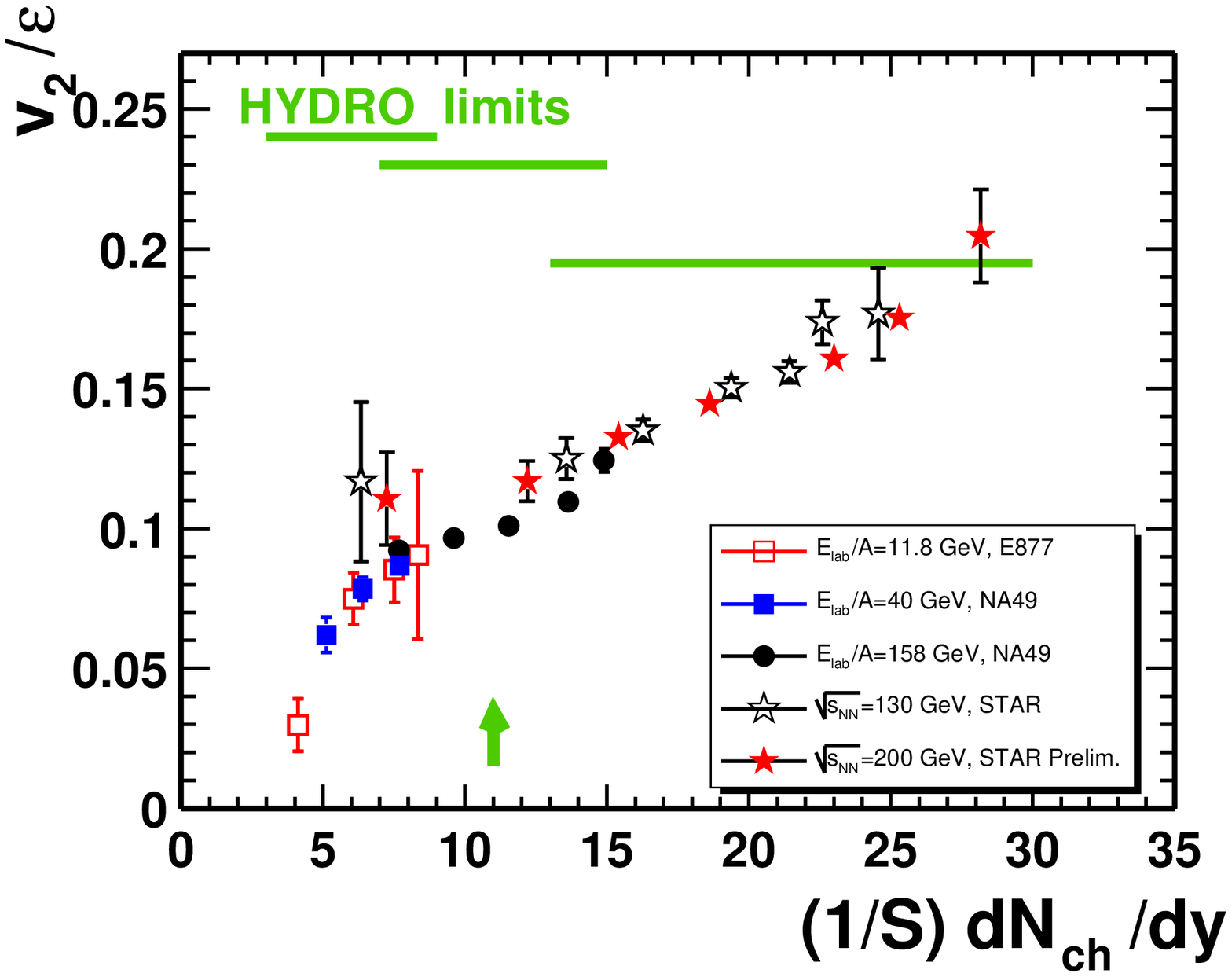}
\begin{minipage}[hbt]{6in}
\caption{{\small First line of evidence: Bulk collective 
flow is the barometric signature
of QGP production. Left figure combines 
STAR~\protect{\cite{Adams:2003zg}-\cite{Adler:2002pu}}
and PHENIX~\protect{\cite{Adler:2003kt}} 
measurements of the azimuthal 
elliptic flow ($v_2(p_T)$) of $\pi,K,p,\Lambda$ in Au+Au at 200 AGeV. 
The predicted hydrodynamic flow pattern 
from \protect{\cite{Kolb:2000fh}-\cite{Kolb:2003dz}} 
agrees well with observations
in the bulk $p_T<1$ GeV domain. Right figure from \protect{\cite{Alt:2003ab}}
shows $v_2$ scaled to the
initial elliptic spatial anisotropy, $\epsilon$, as a function
of the charge particle density per unit transverse area. 
The bulk hydrodynamic limit
is only attained at RHIC.}
\label{line1fig}
}
\end{minipage}
\end{figure}

Figure (\ref{line1fig}) shows the striking bulk collectivity 
elliptic flow signature of QGP formation at RHIC. 
Unlike at SPS and lower energies, the observed large
elliptic deformation ($(1+2 v_2)/(1-2 v_2)\sim 1.5$)
of the final transverse momentum distribution 
agrees for the first time with 
non-viscous hydrodynamic predictions~\cite{Kolb:2000fh}-\cite{Hirano:2004rs}
at least up to about $p_T\sim 1$ GeV/c.
However, the right panel shows that when the local rapidity density per unit 
area~\cite{Voloshin:1999gs,Alt:2003ab} drops 
below the values achieved at RHIC $\sim 30/{\rm fm}^2$, 
then the elliptic flow (scaled by the initial 
spatial ellipticity, 
$\epsilon=\langle (y^2-x^2)/(y^2+x^2)\rangle$) falls below the 
perfect fluid hydrodynamic predictions. 
We will discuss in more detail the origin of
the large discrepancy at SPS energies in the next section. 

The most impressive feature in Fig.(\ref{line1fig})
is the agreement of the observed hadron mass dependence 
of the elliptic flow pattern for 
all hadron species, $\pi, K, p,\Lambda$, with the hydrodynamic predictions
 below 1 GeV/c. This is the QGP fingerprint that shows that there is a common 
 bulk collective azimuthally 
asymmetric flow velocity field, $u^\mu(\tau,r,\phi)$.
Such good agreement with the hadron mass dependence of the
$v_2(p_T,m_h)$ data is furthermore only found when
the input equation of state has the characteristic ``softest point'' near $T_c$
as predicted by lattice QCD~\cite{Kolb:2000fh}-\cite{Hirano:2004rs}.
When equations of state without the predicted drop of  speed of sound near $T_c$
were used as input, the flow velocity field,  especially 
that of the heavy baryon,
was over estimated.

The flow velocity and temperature fields of a perfect (non-viscous)
fluid obeys the  hydrodynamic equations:
\begin{equation}
\partial_\mu\left\{[\epsilon_{QCD}(T(x))+P_{QCD}(T(x))]
u^\mu(x)u^\nu(x)-g^{\mu\nu} 
P_{QCD}(T(x))\right\} = 0 \; ,
\label{hydro}
\end{equation}
where  $T(x)$ is the local temperature field, 
$P_{QCD}(T)$ is the QGP equation of state,
and $\epsilon_{QCD}(T)=(TdP/dT -P)_{QCD}$ is the local proper energy density.
The above equations apply in the rapidity window $|y|<1$,
where the baryon chemical potential 
can be neglected. 
Eq.(\ref{hydro}) provides the barometric
connection between the observed flow velocity 
and the theoretical properties of the QGP.
\begin{figure}[h]
\centering
\includegraphics[height=0.3\textheight,
width=0.45\textwidth,clip]{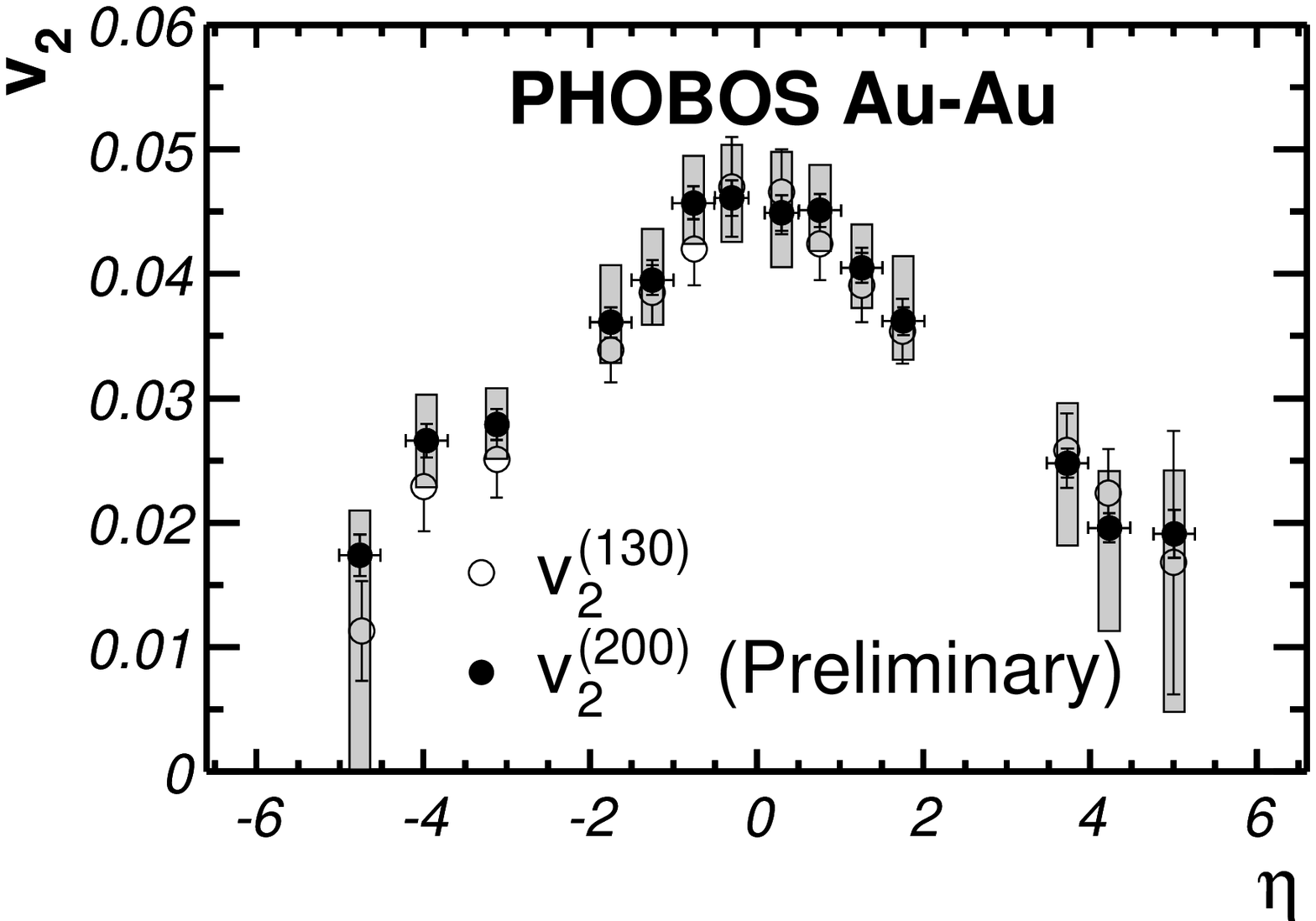}
\includegraphics[height=0.3\textheight,
width=0.45\textwidth,clip]{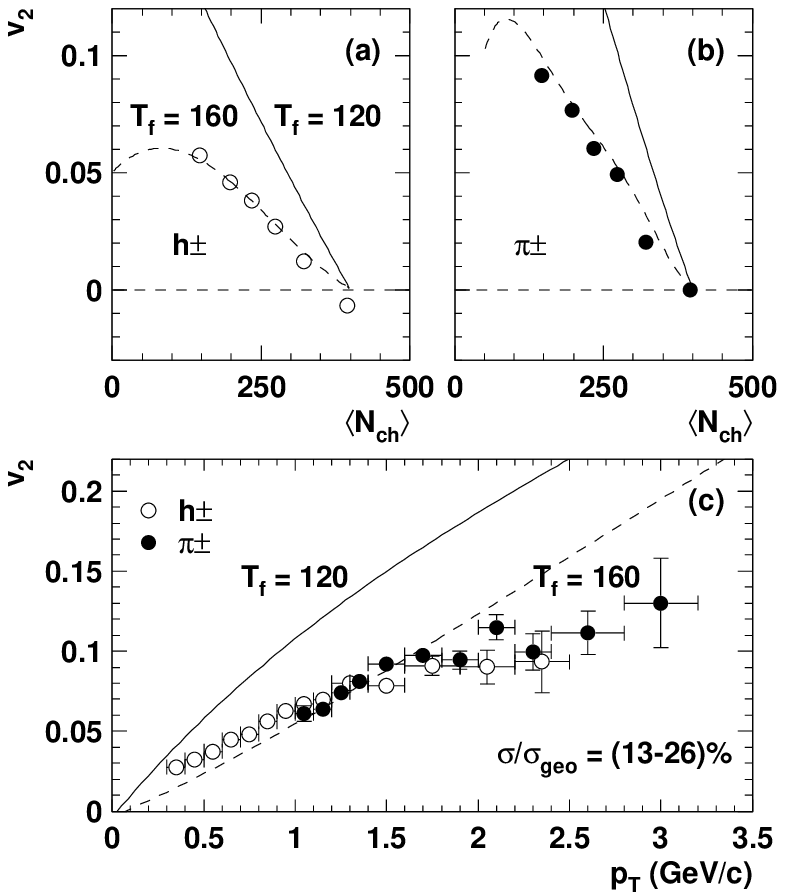}
\begin{minipage}[hbt]{6in}
\caption{  {\small Left figure shows the pseudo rapidity dependence
of elliptic from PHOBOS~\protect{\cite{Back:2002ft}}.
Right figure is CERES{\protect\cite{Agakichiev:2003gg}} 
data on elliptic flow at SPS.  It is well below hydrodynamic predictions 
with freeze-out $T_f=120$ MeV required to reproduce the single inclusive
radial flow. Early freeze-out with $T_f=160$ MeV, simulating effects of
dissipation, is needed to reproduce the data.
}
\label{line1figb}
}\end{minipage}
\end{figure}

We note that in any hydrodynamic treatment of collective flow
there is a tradeoff between the combined effects due to the initial
state boundary condition, the equation of state of the matter, and dissipation
effects. In order to use the flow pattern to constrain the equation of state
the initial condition must be constrained from other measurements
and dissipation must be negligible. Here is where
 the measurements of the global multiplicity systematics\cite{Back:2001ae,dndyphobos} 
are so important. The remarkably weak energy and centrality dependence
of the bulk entropy observed via the $dNdy$ plays a pivotal role as perhaps
the most convincing test of the CGC initial condition hypothesis
 at RHIC\cite{Iancu:2003xm}.
Without such an experimental and theoretical  constraint on the initial condition
no meaningful constraint on the QGP equation of state could have been found.

The study of the interplay between the equation of state and 
dissipative phenomena is more difficult and can only be untangled through detailed systematics
of the flow pattern as a function of beam energy, centrality, and rapidity
dependence.
Here the detailed systematics from AGS and SPS
data have played a pivotal role in helping sorting out the different
viscous effects in hadronic and QGP matter  as we discuss in the next section. 

Why is $v_2$ more emphasized than $v_1$ or radial flow 
as a signature of QGP formation?
The primary reason is that elliptic flow is generated mainly during the 
highest density phase of the evolution before 
the initial geometric spatial asymmetry of the plasma
disappears. It comes from the azimuthal dependence of the pressure gradients,
which can be studied by varying the 
centrality of the events~\cite{Ollitrault:bk}. 
Detailed parton transport~\cite{Molnar:2001ux} 
and hydrodynamic~\cite{Teaney:2001av}
calculations show that most of the 
$v_2$ at RHIC is produced before 3 fm/c and that elliptic flow is 
relatively insensitive to the
late stage dissipative expansion of the hadronic phase.
The reason for the generation of $v_2$ at relatively early times
is that it is very difficult to convert the spatial anisotropy of the matter
distribution into a momentum space anisotropy once the system cools into
the mixed phase, since in the mixed phase pressure gradients cannot
be set up.  Actually, it was a surprise how well 
 the observed collective flow agrees with
perfect fluid
hydrodynamic predictions. Ideal fluid flow requires very strong interactions of 
the quarks and gluons in the plasma  at  early times
$\tau\gton \tau_{eq}\approx 0.6$ fm/c.

In contrast, radial flow has been observed at all energies~\cite{Cheng:2003as}
 and has been shown to be mainly sensitive to late time 
``pion wind'' radial pressure gradients~\cite{Bass:1998vz,Bass:2000ib}, which continue
to blow long after the QGP  condenses into hadronic resonances. 

\subsection{The Breakdown of Bulk Collectivity}

It is important to point out 
that no detailed 3+1D hydrodynamic 
calculation~\cite{Hirano:2003hq}-\cite{Hirano:2004rs}
has yet been able to reproduce the 
rapid decrease of $v_2(|\eta|>1)$ observed by  
PHOBOS in Fig.(\ref{line1figb}). 
This is most likely  due to the 
increasing role  {\em hadronic} dissipation 
effects in the ``corona'' when the comoving density
decreases with increasing $y$. The volume of the QGP shrinks while
the hadronic corona thickens as the rapidity density $dN/dy$ is reduced
within a fixed nuclear geometry away from midrapidity.  
From the right panel of Fig.(\ref{line1fig}), 
we see that  a decrease of the local
transverse density from midrapidity RHIC conditions 
leads to an increasing deviation from the perfect fluid limit.
The initial density was also observed to decrease 
at RHIC as $|y|$ increases~\cite{Bearden:2001qq}.
Therefore, from the known SPS data, we should expect deviations from the 
perfect fluid limit away from the midrapidity region.

Another set of RHIC data that show  deviations
from perfect fluid hydrodynamic predictions is the centrality dependence
of $v_2$.  The observed $v_2(b)$ decreases relative to hydrodynamic
predictions also when the
impact parameter increases toward the more peripheral
collisions. 
This is again due to the fact that the produced multiplicity, $dN/dy\propto N_p(b)$,
decreases with increasing $b$. The hadronization time decreases with $b$
since the QGP is formed with smaller initial density and the hadronic fluid 
is less efficient in transferring pressure into collective flow.

To elaborate further on this important point, 
Fig.\ref{line1figb} shows CERES data~\cite{Agakichiev:2003gg} on
 $v_2(p_T)$  at SPS energy $\sqrt{s}=17$ AGeV. 
In agreement with the NA49 data shown in the right panel
of Fig.(\ref{line1fig}), the CERES data falls well below the hydrodynamic
predictions. At even lower energies, AGS and BEVALAC,
the $v_2$ even  becomes {\em negative} and this 
``squeeze out'' of plane~\cite{Stocker:ci}
 is now well understood in terms of low energy non-equilibrium 
nuclear transport theory~\cite{Stoicea:2004kp,Danielewicz:2002pu}.
  
In order to account for the smallness of $v_2$ at SPS, 
hydrodynamics has to be frozen
out at unphysically high densities and temperatures,
$T_f\approx T_c$. However,
the observed radial flow rules out
this simple fix. The reduction of $v_2$ while maintaining radial flow 
can be approximately
understood in approaches~\cite{ Teaney:2001av} 
that combined perfect fluid
QGP hydrodynamics with dissipative final state hadronic evolution.

In light of the above discussion on the breakdown of collectivity due to hadronic dissipation
at high rapidity and large
impact parameters at RHIC and even at midrapidity at SPS and lower energies,
the smallness of dissipative corrections in the central
regions of RHIC is even  more  surprising.
At mid-rapidities, the lack of substantial dissipation in the QGP phase 
is in itself a remarkable and unexpected discovery \cite{Brown:2004jy} at RHIC.
Calculations based on parton transport
theory~\cite{Molnar:2001ux} predicted large deviations from the ideal
non-viscous hydrodynamic limit even in a QGP.  Instead, the data show that
the QGP at RHIC is almost a perfect fluid. 
A Navier Stokes analysis~\cite{Teaney:2003pb} of the RHIC data
also  indicates  that the 
viscosity of the QGP must be about ten times smaller than 
expected if the QGP were a weakly interactive conventional Debye screened plasma.
This unexpected feature of the QGP must be due to 
strong coupling QCD physics that persists to at least $3T_c$.
(See \cite{Gyulassy:2004vg,Brown:2004jy,Danielewicz:ww} 
and refs therein for further discussion).
\vspace{0.2in}

{\bf Summarizing this section: Elliptic flow measurements confirm that the
quark-gluon matter produced at RHIC is to a very good 
approximation in local thermal equilibrium up to about $3$ fm/c.  
In addition, the final hadron mass dependence of the flow pattern is remarkably
consistent with numerical QCD computations of the equation
of state.  Viscous corrections furthermore appear to be surprisingly small
during this early evolution.
The 
produced Quark Gluon Plasma must therefore be very strongly
interacting.  Such behavior was not seen at lower energy because
the highly dissipative hadronic fluid component masked the QGP flow 
signals. The perfect fluid behavior is also masked at RHIC at higher rapidities
and in more peripheral reactions again due to the increased role 
of the dissipative hadronic ``corona''.}

\subsection{ Perturbative QCD  and Jet Quenching}

In addition to the breakdown of perfect fluid collectivity at high rapidity
seen in Fig.(\ref{line1figb}), 
Fig.(\ref{line1fig}) clearly shows 
that hydrodynamics also breaks down at very short wavelengths and
high transverse momenta, $p_T> 2$ GeV. Instead of continuing 
to rise with $p_T$,
the elliptic asymmetry stops growing and the 
difference between baryon vs meson 
$v_2$ even reverses sign.  Between $2<p_T<5$ GeV the baryon $v_2^B(p_T)$ exceeds the
meson $v_2^M(p_T)$ by approximately 3/2. 
For such short wavelength components of the QGP, 
local equilibrium simply cannot be
maintained due the fundamental asymptotic freedom property
of QCD, i.e. the coupling strength becomes too weak.

In this section, we concentrate on the $p_T>2$ GeV meson observables 
that can be readily understood in terms of QGP modified
perturbative QCD (pQCD) dynamics~\cite{Gyulassy:2003mc,Baier:2000mf}. 
(Baryons at intermediate $2 ~GeV \le p_T \le 5 ~GeV$ are outside the
range of a perturbative treatment and several competing mechanisms
have been proposed and are under theoretical 
development\cite{Kharzeev:1996sq,Csizmadia:1998vp, bmuller}.)

The quantitative study of short wavelength partonic pQCD dynamics 
focuses on the rare high $p_T$
power law tails that extend far beyond the typical (long wavelength) 
scales $p<  3 T \sim 1$ GeV
of the bulk QGP. The second major discovery at RHIC
is that the non-equilibrium power law 
high $p_T$ jet distributions remain power law like 
but are strongly quenched~\cite{Adcox:2001jp}-\cite{Adler:2002ct}.
Furthermore, the quenching pattern has a distinct 
centrality, $p_T$, azimuthal angle,
and hadron flavor dependence that can be used to test the 
underlying dynamics in many independent ways.

\begin{figure}[htp]
\centering
\includegraphics[height=0.3\textheight,width=0.47\textwidth,clip]
{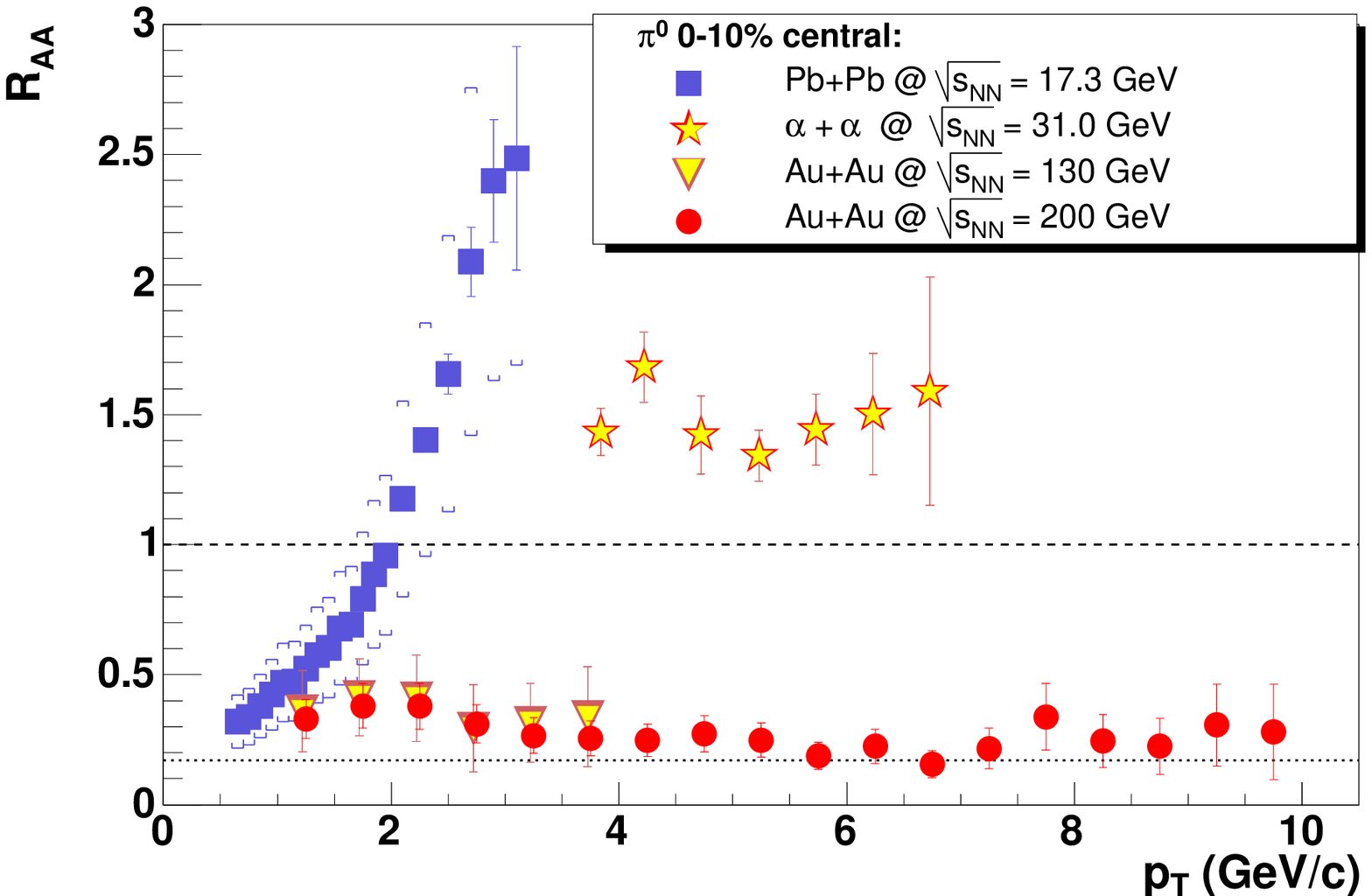}
\includegraphics[height=0.31\textheight,width=0.5\textwidth,angle=-90,
origin=c,clip] {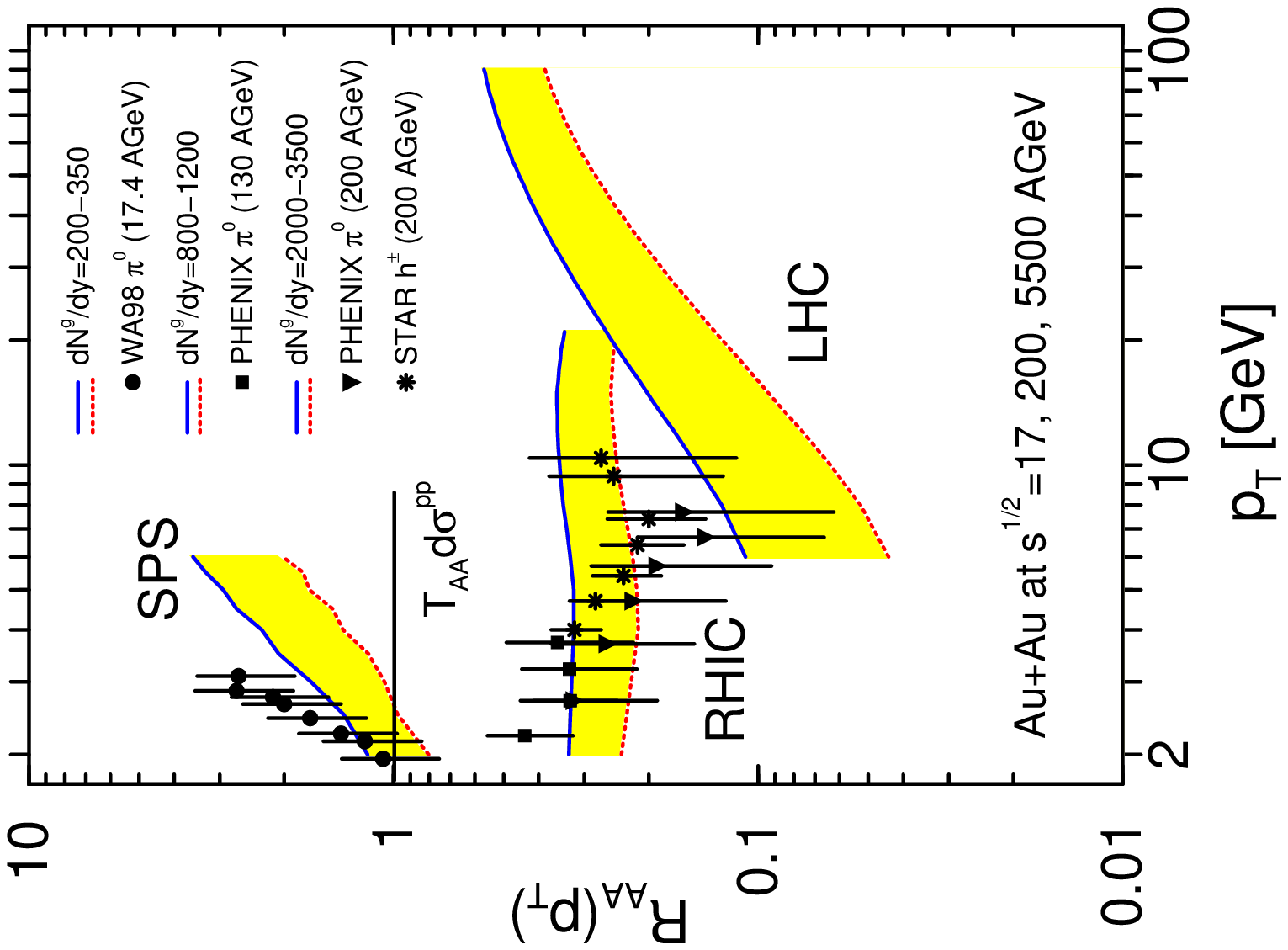}
\begin{minipage}[hbt]{6in}
\caption{{\small Jet Quenching at RHIC.
Left \protect{\cite{d'Enterria:2004ne}} shows the 
jet quenching pattern of $\pi^0$ 
discovered by PHENIX~\protect{\cite{Adcox:2001jp,Adcox:2002pe}}
 at RHIC compared to previous observation of high $p_T$ enhancement
at ISR and SPS energies
  The nuclear modification factor
$R_{AA}= dN_{AA}/T_{AA}(b)d \sigma_{pp}$ measures the deviation of $AA$ spectra
from factorized pQCD. Right shows predictions~\protect{\cite{Vitev:2002pf}} 
of the $\sqrt{s}$
and $p_T$ dependence from SPS, RHIC, LHC based on the GLV 
theory~\protect{\cite{Gyulassy:2000er}} of radiative energy loss.}
\label{line2fig}}
\end{minipage}
\end{figure}

Below RHIC energies, there is an enhancement of moderately
high $p_T$ tails that was observed in central $Pb+Pb$ reactions at the 
SPS.
(Very recent reanalysis of the WA98 data
shows a somewhat weaker enhancement at SPS \cite{d'Enterria:2004ig})
This enhancement was expected as a consequence of the Cronin enhancement: now understood
as an initial
state effect\cite{Accardi:2002ik}t which is also seen in $pA$ collisions.  Since the Cronin
enhancement is an effect of the initial state nuclear wavefunction, it plays
a role in the Color Glass Condensate, but we wish to isolate final state
effects for our study of the Quark Gluon Plasma.
In contrast, at RHIC a 
large suppression, by a factor of 4-5, 
was discovered in central $Au+Au$ that extends 
beyond 10~GeV for $\pi^0$.  

Jet quenching in $A+A$ was proposed in~\cite{Gyulassy:1990bh,Wang:1992xy} 
as a way to study the dense matter produced at RHIC energies. 
As noted before, the pQCD 
jet production  rates finally become large enough to measure yields up to 
high $p_T > 10$ GeV.
Order of magnitude suppression effects were predicted based on simple
estimates of induced gluon radiative energy loss.
Ordinary, elastic energy loss~\cite{Bjorken:1982tu} 
was known by that time to be too small to lead to significant attenuation.

As reviewed in~\cite{Gyulassy:2003mc,Baier:2000mf}
refinements in the theory since then have opened the possibility of 
using the observed jet quenching pattern
as a tomographic tool~\cite{TOMO} 
to probe the parton densities in a QGP. The right panel of Fig.\ref{line2fig} shows
a recent jet tomographic analysis~\cite{Vitev:2002pf} of the PHENIX 
$\pi^0$ data~\cite{Adcox:2001jp,Adcox:2002pe}
based on the GLV opacity formalism~\cite{Gyulassy:2000er}.
This analysis concludes that the 
initial gluon rapidity density 
required to account for the observed jet quenching pattern must be
$dN_g/dy\sim 1000\pm 200$. 

This jet tomographic measure of the initial $dN_g/dy$ 
is in remarkable agreement with three other independent
sources: (1) the initial  entropy
deduced via the Bjorken formula from the measured multiplicity, (2) 
the initial condition of the QGP required in hydrodynamics 
to produce the observed elliptic flow, 
and  (3) the 
estimate of the maximum gluon rapidity density
bound from the CGC gluon saturated initial condition, (which will be
described later).

These four independent measures make it possible to estimate 
the maximal initial energy density in central collisions 
\begin{equation}
\epsilon_0 = \epsilon(\tau \sim 1/p_0) 
\approx \frac{p_0^2}{\pi R^2}\frac{dN_g}{dy}
\approx 20 \frac{{\rm GeV}}{{\rm fm}^3} \sim 100 \times \epsilon_{A}
\label{eps0}
\end{equation}
where $p_0\approx Q_{sat}\approx 1.0-1.4$ GeV is the mean transverse momentum
of the initial produced gluons from the incident saturated virtual nuclear
CGC fields\cite{McLerran:2004fg,Iancu:2003xm,Eskola:2002qz}. 
This scale controls the
formation time $\hbar/p_0\approx 0.2$ fm/c of the initially out-of-equilibrium 
(mostly gluonic) QGP.  
The success of the hydrodynamics
requires that local equilibrium  be achieved on a 
fast proper time scale $\tau_{eq}\approx (1-3)/p_0 < 0.6$ fm/c.
The temperature at that time is 
$T(\tau_{eq})\approx (\epsilon_0/(1-3)\times 12)^{1/4} \approx 2 T_c$.

In HIJING model\cite{ToporPop:2002gf}, 
the mini-jet cutoff is $p_0=2-2.2$ GeV  limits the number of mini-jets
to well below 1000. The inferred opacity of the QGP is observed to be much higher 
and consistent with 
the CGC\cite{Iancu:2003xm} and EKRT\cite{Eskola:2002qz} estimates.

\subsubsection{\protect{$I_{AA}$} and Di-Jet Tomography}  

\begin{figure}[htp]
\centering
\vspace*{0.4cm}
\includegraphics[height=0.33\textheight,width=0.4\textwidth,clip] 
{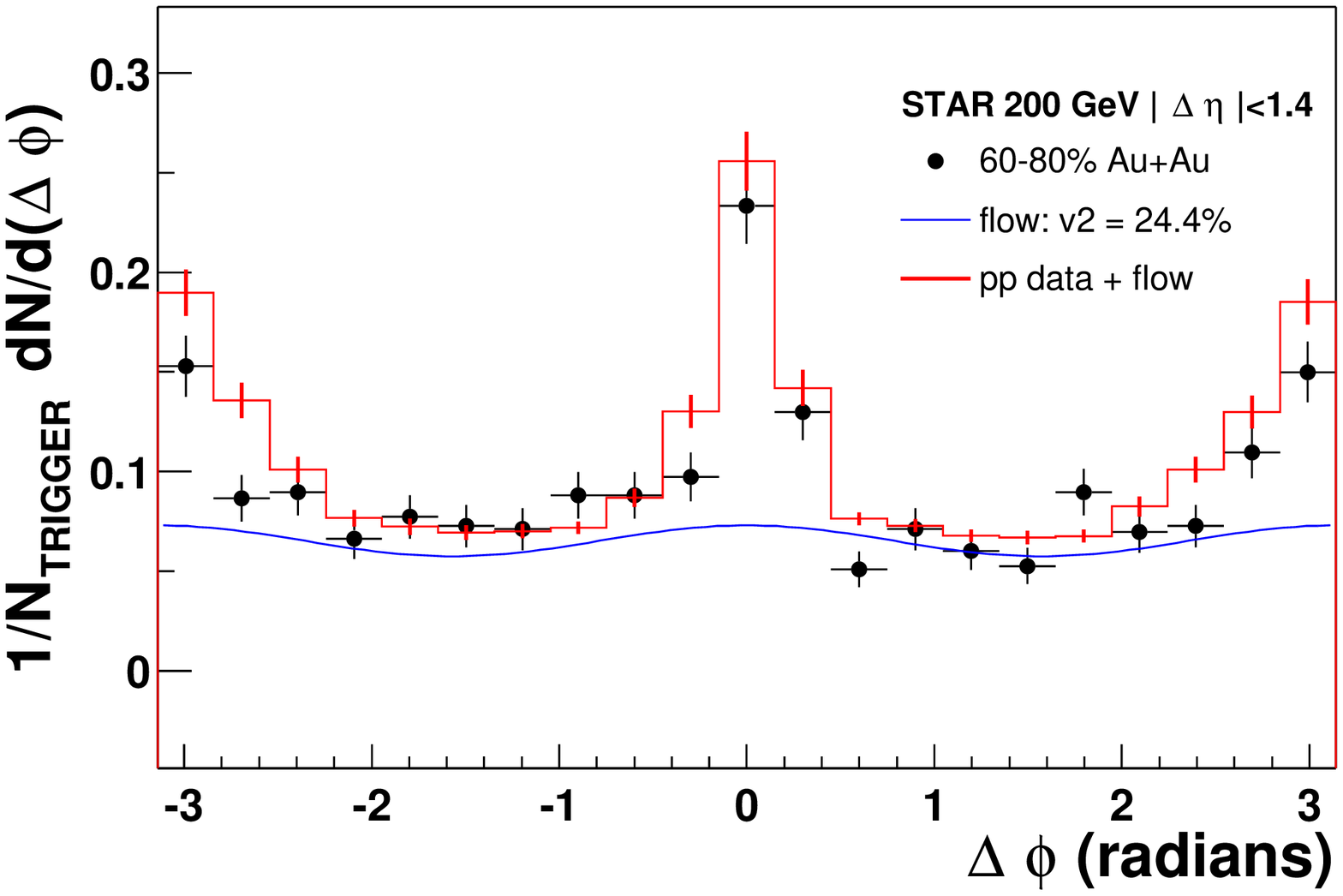}
\includegraphics[height=0.33\textheight,width=0.4\textwidth,clip] 
{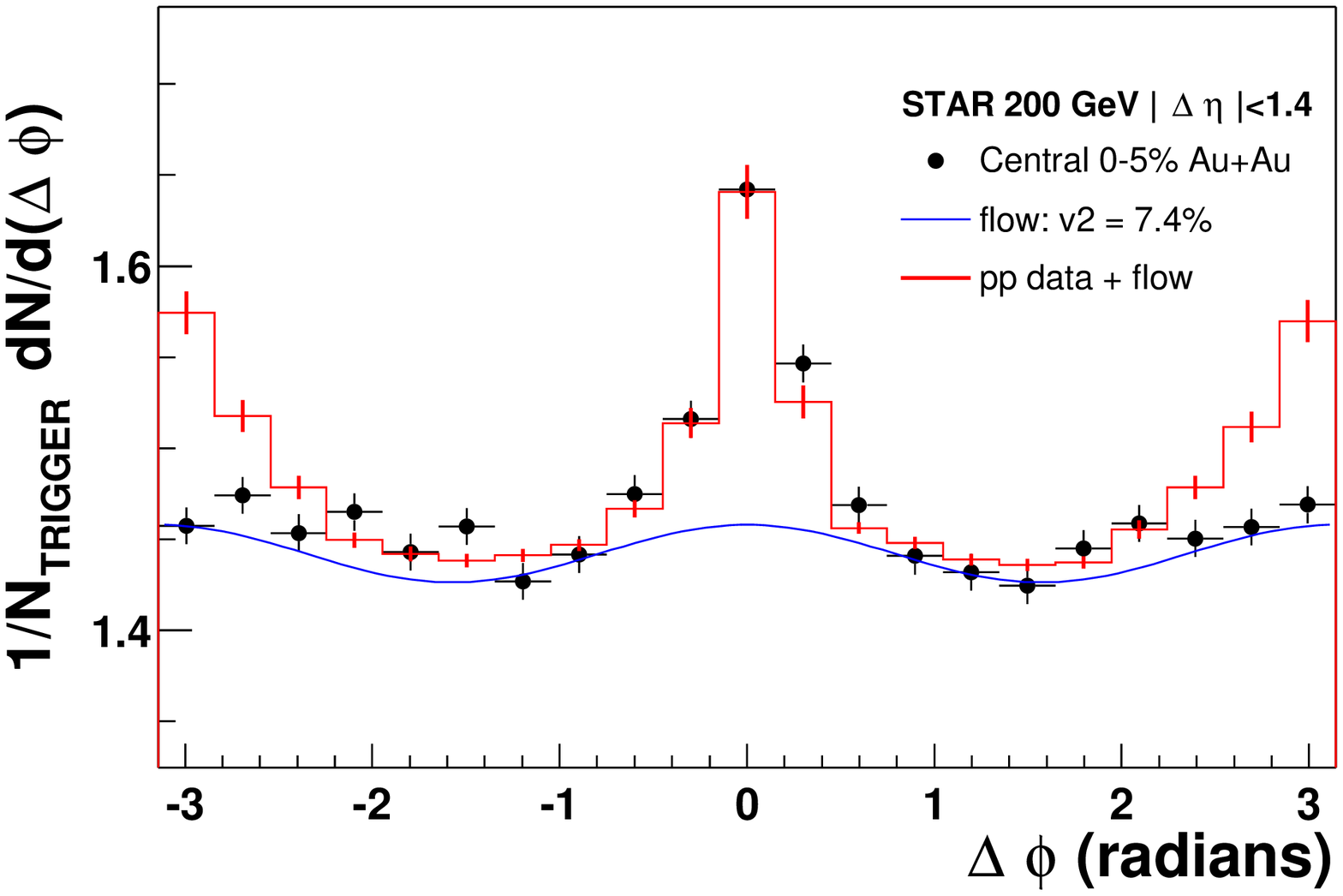}
\begin{minipage}[hbt]{6in}
\caption{{\small Monojets at RHIC
 from STAR~\protect{\cite{Adler:2002tq,Hardtke:2002ph,Jacobs:2003bx}}. 
Strongly correlated back-to-back di-jet production 
in $pp$ and peripheral $AuAu$ left side is compared to mono-jet production
discovered in central $AuAu$.}
\label{monojet}}
\end{minipage}
\end{figure}

Measurements of near side and away side azimuthal angle correlations of
di-jet fragments provide the opportunity to probe
the evolution of the matter produced at RHIC in even more detail.
Fig.(\ref{monojet}) show the 
discovery~\cite{Jacobs:2003bx,Adler:2002tq,Hardtke:2002ph} 
of mono-jet production~\cite{Gyulassy:1990bh} in central 
collisions at RHIC.
In peripheral collisions, the distribution $dN/ d\Delta \phi$ of the azimuthal
distribution of $p_T\sim 2$ GeV hadrons relative to a tagged $p_T\sim 4$ GeV
leading jet fragment shows the same 
near side and away side back-to-back jetty correlations
as measured in $p+p$. This is strong evidence that the kinematic range studied
tests the physics of pQCD binary parton collision processes. 
For central collisions, on the 
other hand, away side jet correlations are almost completely
suppressed.

The published data are as yet limited to $y_1\approx y_2\approx 0$, broad $p_T$
cuts: $p_{T1} > 4$ GeV and $p_T\sim 2$ GeV, 
two bins of $\phi_1-\phi_2$, and of course averaged over $\Phi_b$.
The measured modification of di-jet correlations is obtained by subtracting 
out 
the correlations due to bulk elliptic flow, and this introduces some
uncertainty.  Analysis of present and future data at higher transverse
momenta for a variety of rapidities will allow better tests of the underlying
perturbative QCD dynamics.

Only one year ago~\cite{transdyn03} the interpretation of high 
$p_T$ suppression
was under intense debate because it was not yet clear how much of the quenching
was due to initial state saturation (shadowing) of the gluon 
distributions and how much
due to jet quenching discussed in the previous section.
There was only one way to find out - eliminate the QGP final state interactions
by substituting a  Deuterium beam for one of the two heavy nuclei. In fact,
it was long ago anticipated~\cite{Wang:1992xy} 
that such a control test would be needed to isolate
the unknown nuclear gluon shadowing contribution to the A+A quench pattern.
In addition $D+Au$ was required to test predictions of possible  
initial state Cronin multiple 
interactions~\cite{Accardi:2002ik,Wang:1998ww,Wang:1996yf,Vitev:2003xu,Qiu:2003vd}.
In contrast, one  model of 
CGC~\cite{Kharzeev:2002pc} predicted a substantial suppression in $D+Au$
collisions.
The data~\cite{Adler:2003ii,Adams:2003im,Arsene:2003yk,Back:2003ns} 
conclusively
rule out large initial shadowing as the cause of the $x_{BJ} > 0.01$
quenching in Au+Au. 

\begin{figure}[t]
\centering
\includegraphics[height=0.30\textheight,width=.42\textwidth,clip]
{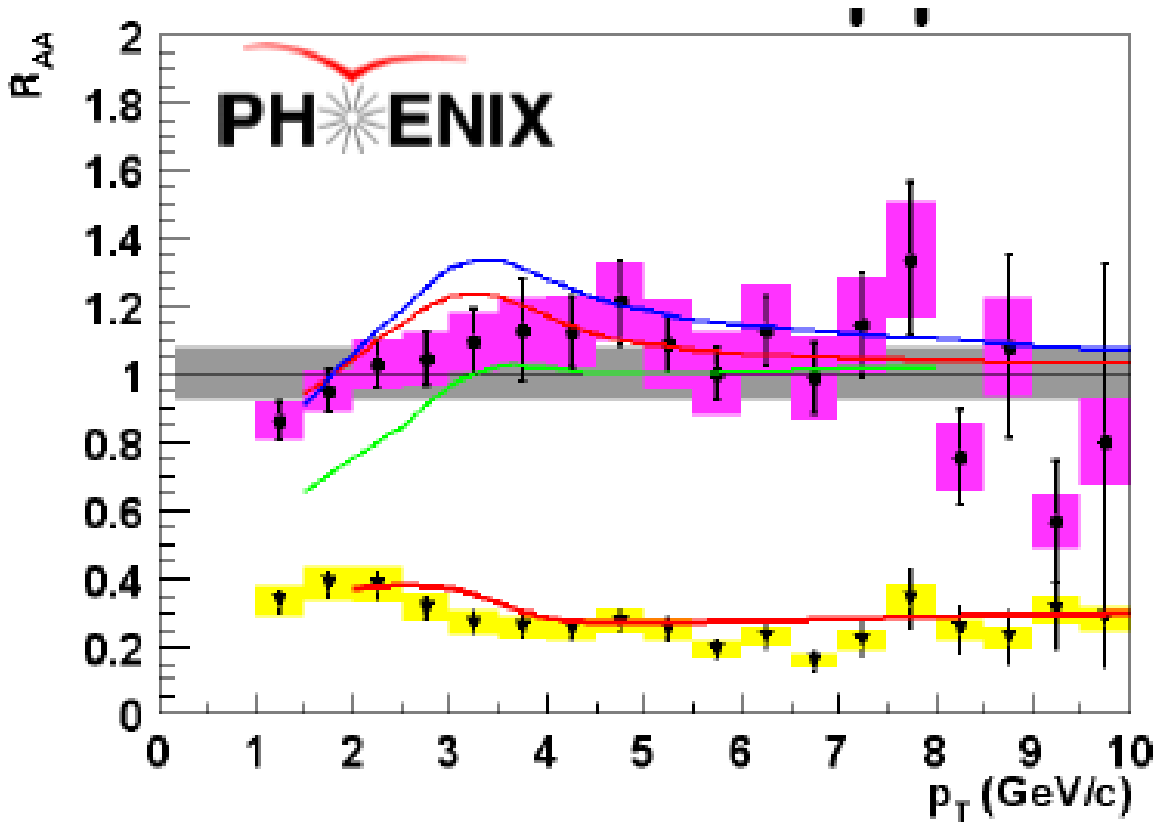}
\includegraphics[height=0.30\textheight,width=.42\textwidth,clip]
{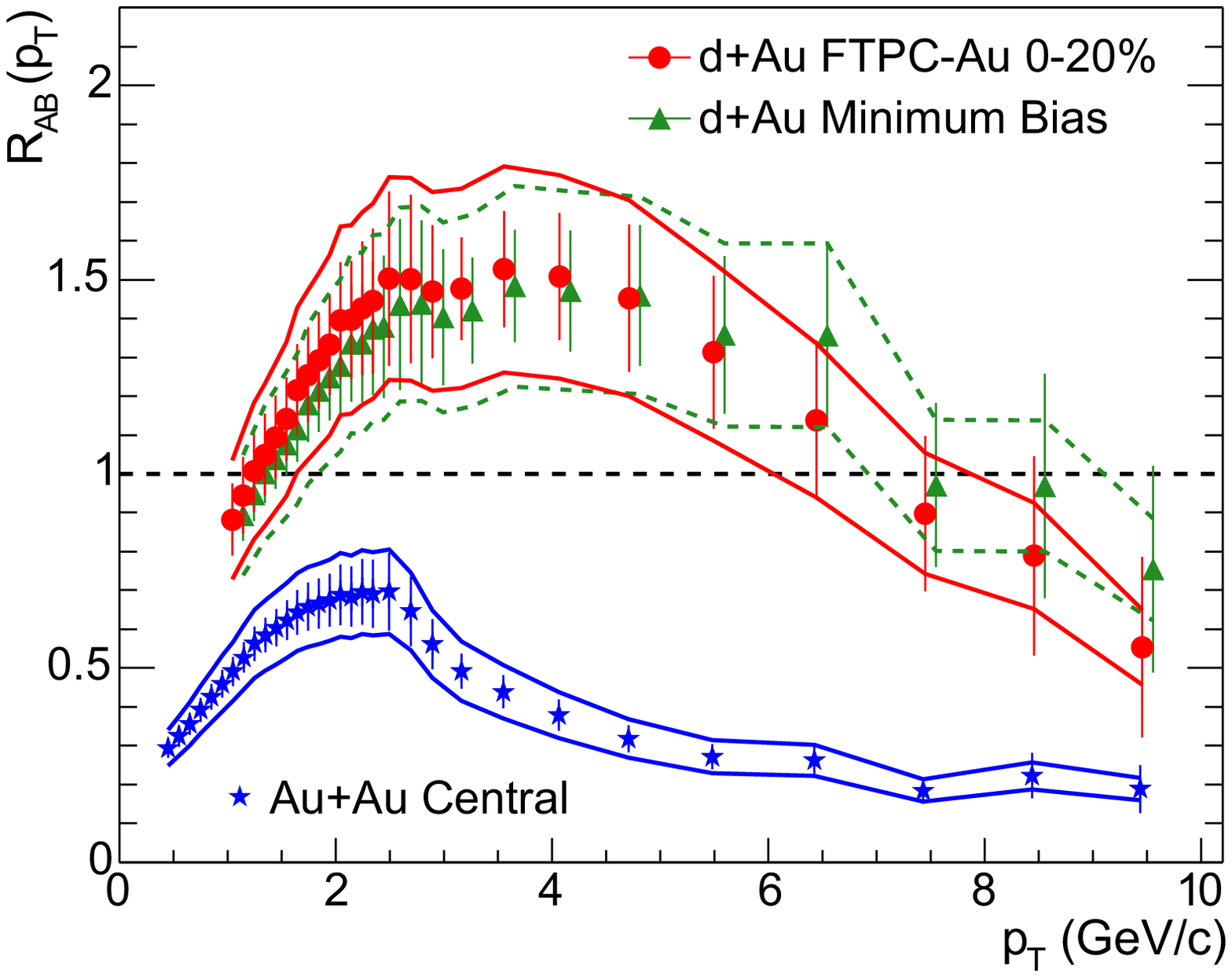}
\begin{minipage}[hbt]{6in}
\caption{{\small The {\bf dA} control: PHENIX~\protect{\cite{Adler:2003ii}} $\pi^0$ 
and STAR~\protect{\cite{Adams:2003im}} $h^{\pm}$
data compare $R_{DAu}$ to $R_{AuAu}$. These and 
BRAHMS~\protect{\cite{Arsene:2003yk}}
and PHOBOS~\protect{\cite{Back:2003ns}} 
data prove that jet quenching in $Au+Au$
must be due to final state interactions. Curves for 
$\pi^0$ show predictions from 
\protect{\cite{Vitev:2002pf}} for $AuAu$ and from
\protect{\cite{Vitev:2003xu}} $DAu$. The curves for $DAu$ show the interplay
between different gluon shadow parameterizations (EKS, none, HIJING) and 
Cronin enhancement
and are  similar to 
predictions in \protect{\cite{Wang:1998ww,Wang:1996yf,Vitev:2003xu}}.
In lower panel, the unquenching of charged hadrons 
is also seen in $D+Au$ relative to $Au+Au$ at high $p_T$.}
\label{rdaudata}}
\end{minipage}
\end{figure}

The $I_{DAu}$ measurement from STAR~\cite{Adams:2003im} shows clearly how
the suppression disappears in $D+Au$ collisions.
\begin{figure}[h]
\centering
\includegraphics[height=0.35\textheight,width=0.8\textwidth,clip]
{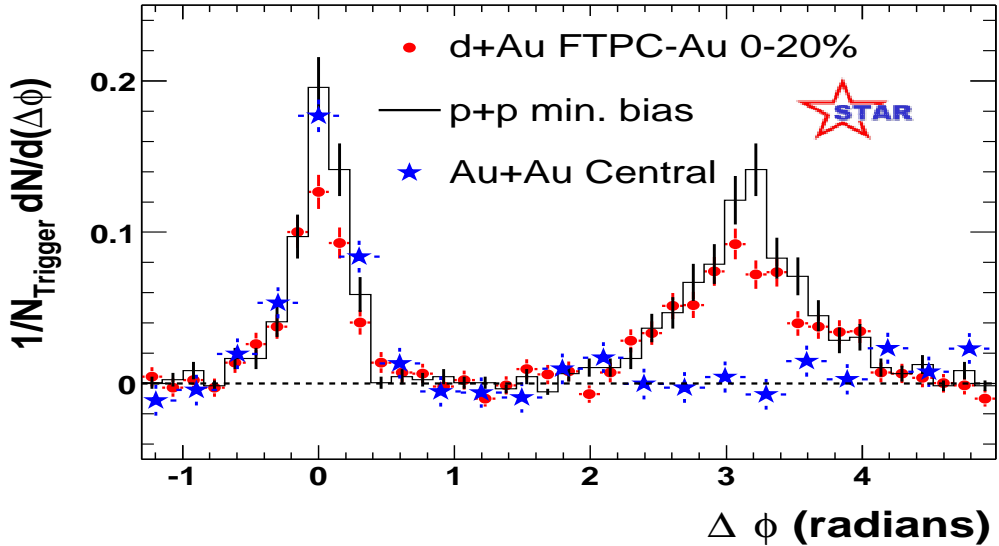}
\begin{minipage}[hbt]{6in}
\caption{{\small The {\bf dA} ``Return of the Jeti'': 
Dijet fragment azimuthal correlations from STAR~\protect{\cite{Adams:2003im}} 
in $DAu$ are unquenched relative to the mono jet correlation
observed in central ${AuAu}$. }
\label{jeti}}
\end{minipage}
\end{figure}
The return of back-to-back jet correlation in $D+Au$ to the level 
observed in $pp$
is seen in Fig.\ref{jeti}. 
The data appear to be entirely consistent with jet 
quenching as a final state effect in $AuAu$ with little initial state
effect in $D+Au$.
These $D+Au$ data support the conclusion~\cite{Wang:2003mm,Wang:2003aw}
that the observed jet quenching in $AuAu$ is due to parton energy loss.

\vspace{0.2in}
{\bf Theoretical analysics of jet quenching confirm the energy density
estimates determined from measurements of particle multiplicity.  
They give large energy losses for jets propagating through the matter
produced at RHIC, and strengthen the case for multliple strong
interactions of the quark and gluon constituents of the matter made at RHIC.}

\section{Empirical Evidence for the Color Glass Condensate}

In this section,  the accumulated evidence
for the Color Glass Condensate hypothesis is discussed\cite{McLerran:2004fg}. 
The evidence rests in a
variety of measurements done at different accelerators with different
type of particles scattering.  The discussion of the results from RHIC will
be emphasized here, {but it is first important to
briefly review the results from HERA involving electron proton
scattering.}

\subsection{Results from Electron-Hadron Scattering}

Electron-hadron scattering provide information about the
wavefunction of a hadron.  The Color Glass Condensate describes the
contribution to this wavefunction which have very many gluons in them.
These pieces of the wavefunction control the physics at very small x,
typically $x \le 10^{-2}$.
The various pieces of experimental information which support the CGC 
hypothesis come largely from $ep$ scattering experiments at HERA:
\begin{figure}[t]
    \begin{center}
        \includegraphics[width=0.50\textwidth]{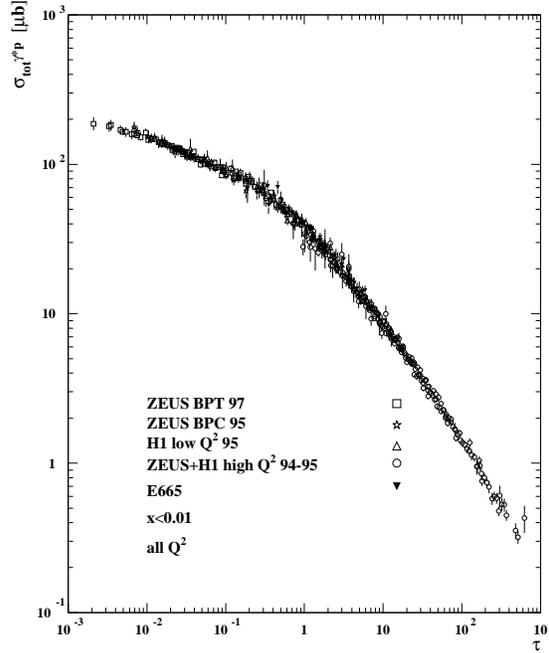}
\begin{minipage}[hbt]{6in}
        \caption{{\small The cross section $\sigma^{\gamma^*p}$ as a function
of the scaling variable $\tau = Q^2/Q_{sat}^2$.\protect{\cite{gbks}}
 }
\label{gb}}
\end{minipage}
    \end{center}
\end{figure}

\begin{itemize}

\item{Geometrical Scaling}

Geometrical scaling is the observation\cite{biel}-\cite{gbks}
that the deep inelastic cross section for virtual photon scattering
as a function of $Q^2$ and $x$ is really only a function of
\be
        \sigma^{\gamma^* p} \sim F(Q^2/Q_{sat}^2)
\ee
where the saturation momentum {increases as the fractional
momentum, $x$, of the gluon tends to zero as}
\be
        Q_{sat}^2(x)  \sim (x_0/x)^\lambda ~ 1 GeV^2
\ee
with $\lambda \sim 0.3$ and $x_0 \sim 10^{-4}$.  This scaling 
{with $\tau=Q^2/Q_{sat}^2$ } works 
for $x \le 10^{-2}$ and over the available  $Q^2$ {range at HERA } 
 as shown in Fig. \ref{gb}.

It is straightforward to understand why this scaling works for the small
$Q^2 \le Q_{sat}^2$.  This is the region of the CGC, and there is only
one dimensionful scale which characterizes the system: the saturation 
momentum.\cite{lt}  The surprise is that there is an extended
scaling window for 
$Q_{sat}^2 \le Q^2 \le Q_{sat}^4/\Lambda^2_{QCD}$.\cite{iim} 
This can be proven analytically.  As well, one now has reliable
computation of the dependence on $x$ of the saturation momentum, that is,
one knows the exponent $\lambda$ to about $15\%$ accuracy, and it agrees
with what is seen from the geometrical scaling curve.\cite{mt}  
What is not determined
from the theory of the CGC is the scale $x_0$, and this must be found by
experiment.  This comes from the boundary conditions for the
renormalization group equations.

\item{The Structure Function $F_2$ }

Using the dipole description of the virtual photon wavefunction,
the structure function $F_2$ can be related to the gluon distribution
function which arises from the CGC.  
The results for the description of the data are remarkably good
for $x \le 10^{-2}$ and $Q^2 \le 45 ~GeV^2$.
One should note that this description includes both the high and low
$Q^2$ data.  Descriptions based on DGLAP evolution can describe the large
$Q^2$ points.  The CGC description is very economical in the number
of parameters which are used \protect\cite{iimu}.

\item{Diffraction and Quasi-Elastic Processes}

The CGC provides a description of the underlying structure of gluonic
matter inside a hadron.  As such, it should be sensitive to probes of 
the transverse extent of this matter, which can be
experimentally studied in  diffraction and related 
quasi-elastic particle production.\cite{strikman}-\cite{gbb}

Diffractive scattering can be computed in the CGC description for
small inclusive masses of produces particles.  The CGC agrees both
with generic features of the data, and provides a reasonably good
quantitative description.

There are additional computations of quasi-elastic $\rho$ meson 
production and $J/\Psi$ production\cite{mms}-\cite{kowt}.  Again,
up to uncertainties associated with the overall normalization (which
arises from imprecise knowledge of hadronic wavefunctions), the CGC 
hypothesis provides a reasonably good quantitative description.

\end{itemize}

\subsection{Heavy Ion Collisions}

The collision of two ultrarelativistic heavy ions can be visualized 
as the scattering of two sheets of colored glass, as shown in 
Fig. \ref{sheetonsheet}. \cite{mkw}-\cite{nkv}

At very early times after the collision the matter is at very high energy
density and in the form of a CGC.  As time goes on, the matter expands.
As it expands the density of gluons decreases, and gluons begin
to propagate with little interaction.
At later times, the interaction
strength increases and there is sufficient time for the matter to
thermalize and form a Quark Gluon Plasma.  This scenario is 
shown in Fig. \ref{times}, with realistic estimates for energy density
and time scales appropriate for the RHIC heavy ion accelerator.

\subsection{The Multiplicity}

\begin{figure}[ht]
    \begin{center}
        \includegraphics[width=0.50\textwidth,angle=270.]
{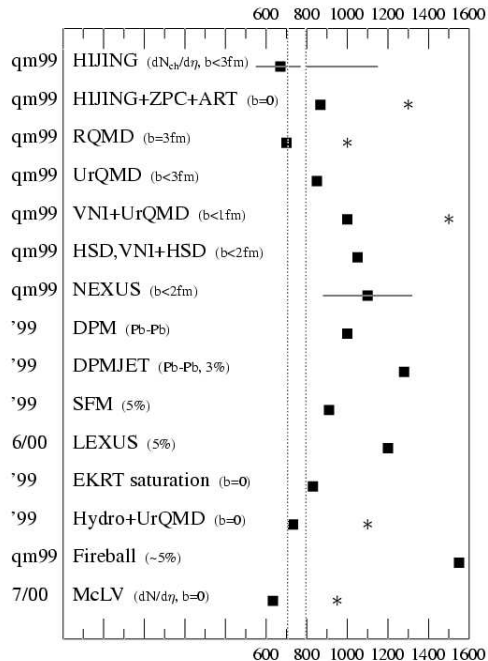}
\begin{minipage}[hbt]{6in}
        \caption{{\small Predictions for the charged hadron rapidity density
for central $Au+Au$ at $\sqrt{s}=200$ AGeV compiled by Eskola \protect\cite{Eskola:2001vs} 
compared to data band ($740\pm 50)$ measured later 
by all four experiments at RHIC. The CGC prediction
 is marked McLV.}
\label{dndypred}}
\end{minipage}
    \end{center}
\end{figure}

The CGC allows for a direct computation of the particle multiplicity
in hadronic collisions.  If one naively tries to compute jet production,
the total multiplicity is infrared divergent.  This follows because of
the $1/p_T^4$ nature of the perturbative formula for gluon production
\be
        {1 \over {\pi R^2}}{{dN} \over {dyd^2p_T}} \sim {1 \over \alpha_S}~
{{Q_{sat}^4} \over p_T^4}
\label{cgchigh}
\ee
In the CGC, when $p_T \le Q_{sat}$, this divergence is cutoff and 
the total gluon multiplicity goes as
\be
        {1 \over {\pi R^2}} {{dN} \over {dy}} \sim {1 \over \alpha_S}~ 
Q_{sat}^2
\label{cgclow}
\ee
One can compute the proportionality constant and before the RHIC data appeared,
predictions were made for the gluon multiplicity.  In Fig. \ref{dndypred} from
Ref.\cite{Eskola:2001vs},
the predictions for 200 AGeV to extrapolations from  the first RHIC run 
at 130 AGeV are shown.  
The CGC correctly predicted the surprizingly low multiplicity.

\begin{figure}[ht]
    \begin{center}
        \includegraphics[width=0.80\textwidth]{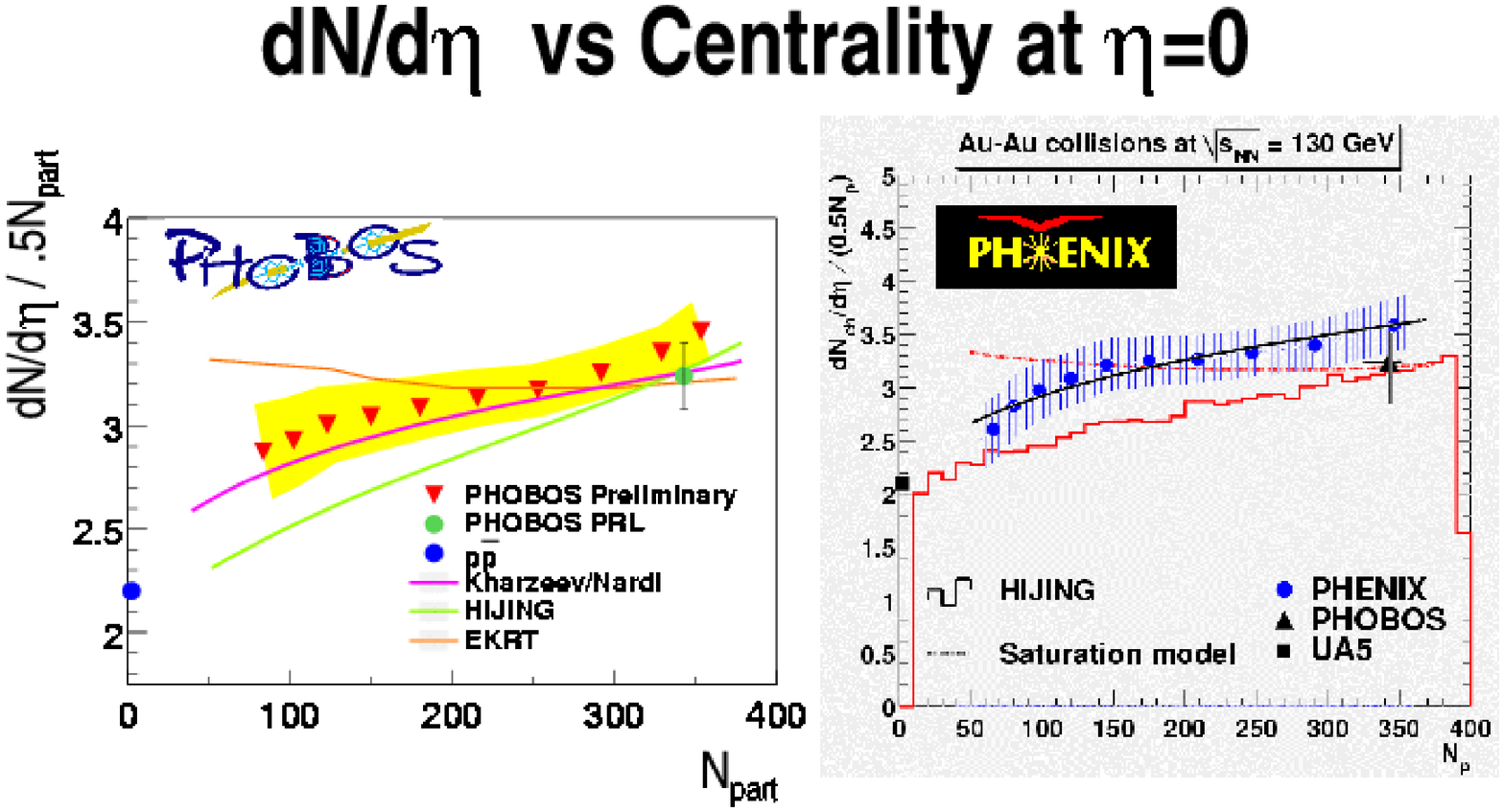}
\begin{minipage}[hbt]{6in}
        \caption{{\small The total multiplicity as a function of the number
of participants as measured by Phobos and Phenix. 
{\protect{\cite{Back:2002ft, Back:2000gw, Adcox:2001mf}} }
 }
\label{glass1}}
\end{minipage}
    \end{center}
\end{figure}

Also, the dependence of the multiplicity on the number of participants
can also be computed, realizing that the saturation momentum squared should be 
(for not too small x) proportional $N_{part}^{1/3}$.  This leads to
\be
        \frac{1}{N_{part}}{{dN} \over{dy}} \sim {1 \over \alpha_S(Q^2_{sat})}
\sim \log N_{part}
\label{cgcdndy}
\ee
so that  a very slow logarithmic dependence on the number of 
participants is predicted in agreement with experiment,
as shown in Fig. \ref{glass1}.\cite{ekrt}-\cite{phenixnpart}

One can go {much} further\cite{Hirano:2004er}
 and compute the dependence of the multiplicity
on rapidity and centrality, and the transverse momentum distribution of 
produced hadrons by using CGC initial conditions matched together with
a hydrodynamic calculation which evolves the matter through the
Quark Gluon Plasma\cite{Hirano:2004rs}. At high $p_T \gton 2$ GeV, jet quenching 
of the pQCD power law tails is also taken into account using \cite{Gyulassy:2000er}.
 This combined
hydrodynamic calculation of Nara and Hirano accounts very well 
for many important features of the data at RHIC.
\cite{Adcox:2001jp},\cite{Back:2000gw}
\begin{figure}[htb]
    \centering
       \mbox{{\epsfig{figure=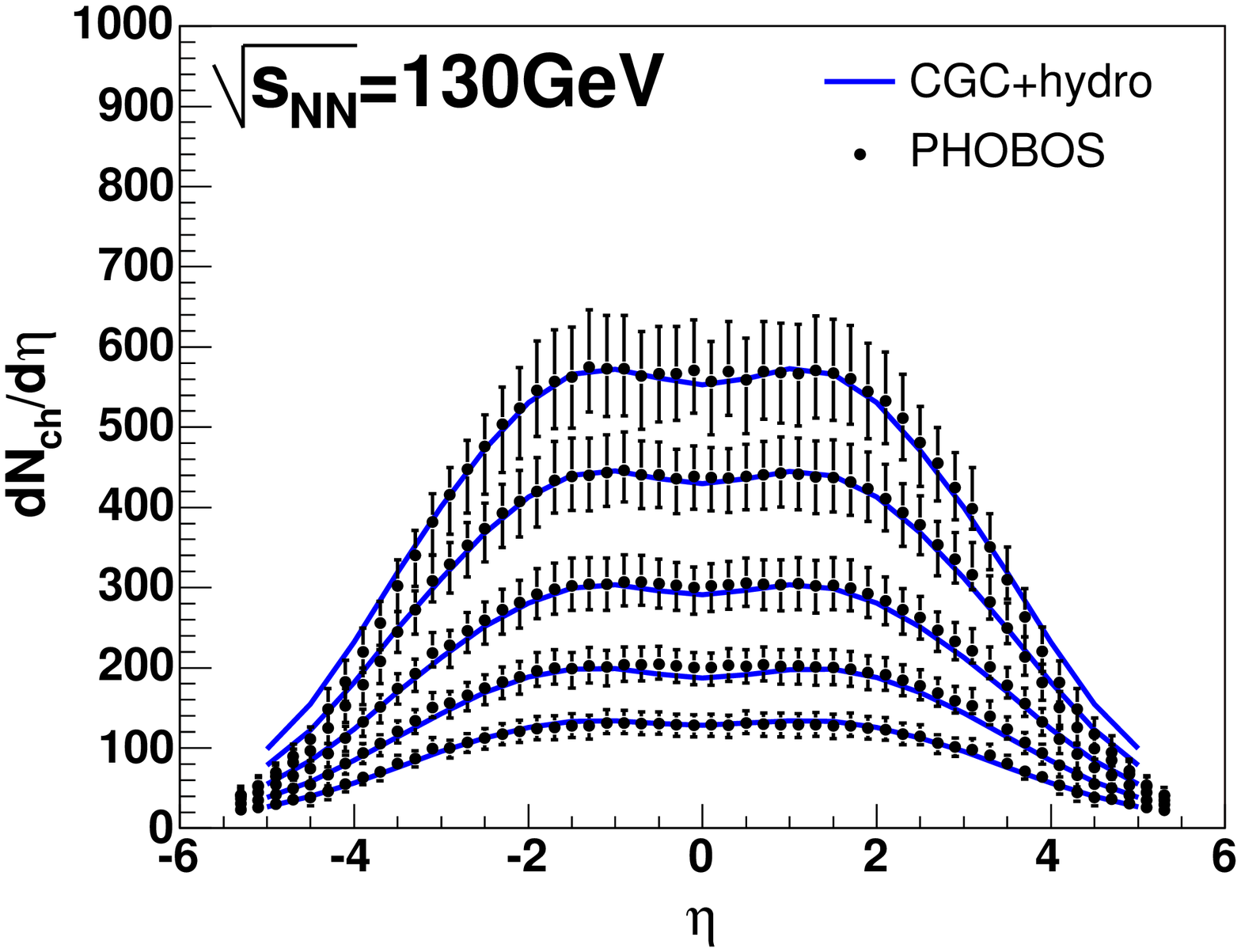,  
        width=0.45\textwidth}}\quad
             {\epsfig{figure=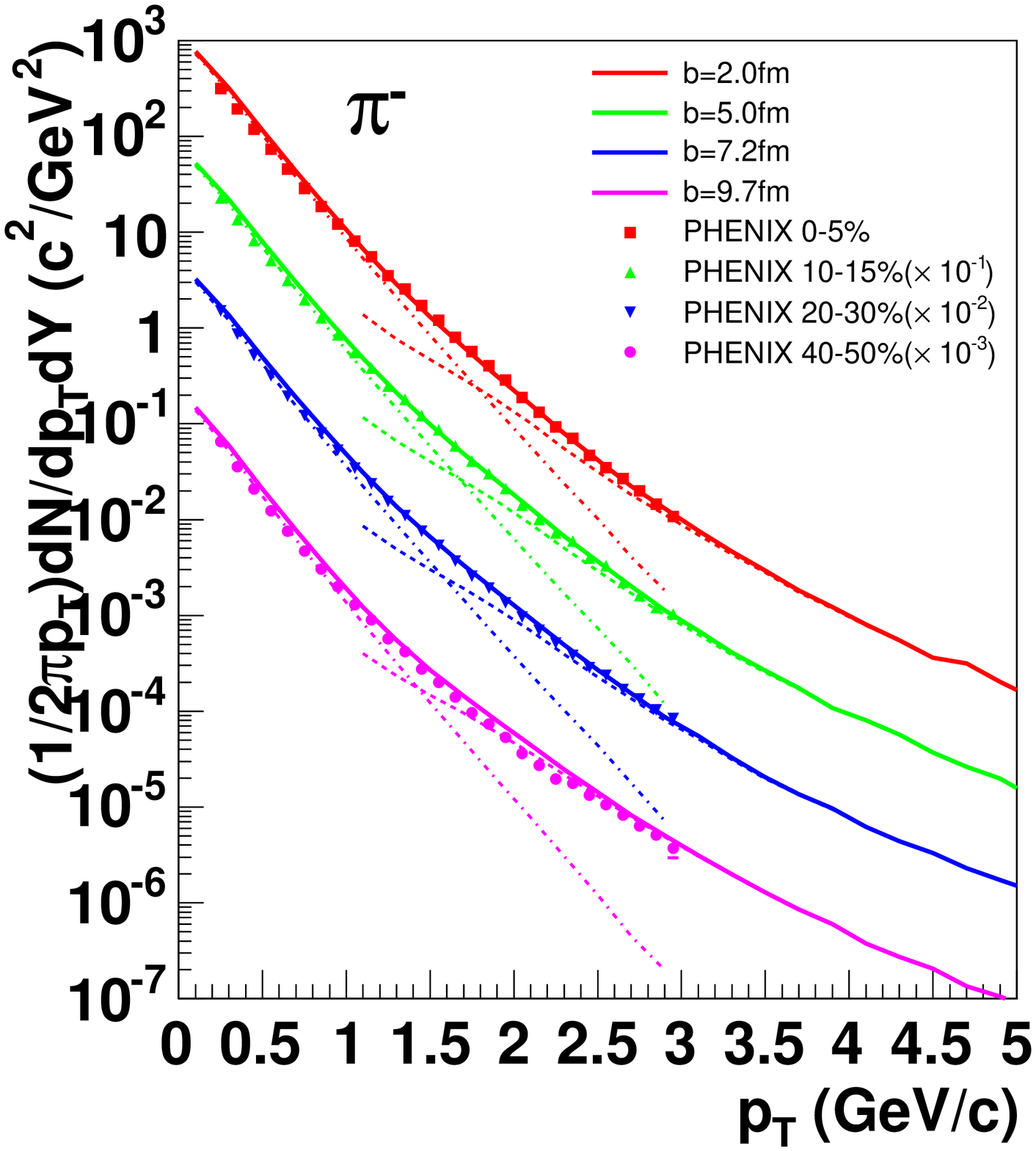,  
        width=0.45\textwidth}}}
\begin{minipage}[hbt]{6in}
        \caption{{\small 
Results of hydrodynamic calculations\protect{\cite{Hirano:2004rs}} with
a CGC initial condition. {Left is the centrality dependence
of the rapidity density from PHOBOS\protect{\cite{Back:2000gw}}. 
 Right shows the transverse momentum distribution at zero
rapidity for different centralities from PHENIX\protect{\cite{Adcox:2001jp}}. 
For $p_T> 2$ GeV jet quenching\protect{\cite{Gyulassy:2000er}} through
the evolving QGP is also taken into account.}}
        \label{hirano}}
\end{minipage}
\end{figure}

We emphasize the importance of constraining the QGP hydrodynamics
with a QCD theoretic initial condition, the CGC. At SPS energies, the uncertainty
about initial conditions together with failure of hydrodynamics to account for
the highly dissipative hadronic flow are in remarkable contrast
to situation at RHIC. If the initial condition were not well constrained at RHIC,
then the conclusion that a QGP was formed could not be sustained.
Hydrodynamics is a dynamical mapping of a given given initial condition to final 
spectra that depends on the equation of state. Only with a known or
predicted initial state does that mapping have the power to falsify 
the QCD equation of state.

In Figs 15 and 16, one can see that several phenomenological models, 
such as HIJING\cite{ToporPop:2002gf}, could also account qualitatively 
for some of the global multiplicity observables.
However, the surprising very weak centrality and beam energy dependence 
observed\cite{Back:2002ft,Bearden:2001qq, Back:2000gw, Adcox:2001mf}
is most satisfactory explained and predicted
by the CGC as arising from the slow $1/\alpha_S\sim \log Q_{sat}^2(N_{part},\surd s)$ in 
eqs.(\ref{cgclow},\ref{cgcdndy}). This is one of the strongest lines 
of empirical evidence from RHIC that
the CGC initial state (with its predicted $N_{part}$ and $\surd s$ dependence)
is formed and that it is the seed of the QGP that evolves from it.

\subsection{High $p_T$ Particles}

The early results from RHIC on gold-gold collisions revealed that the high
$p_T$ production cross sections were almost an order of magnitude below 
that expected for jet production arising from incoherent parton-parton 
scattering.\cite{expjetsuppression} This could be either due to initial state
shadowing of the gluon distribution inside the nuclei,\cite{klm}
or to final state jet 
quenching.\cite{thejetsuppression}  For centrally produced jets, the x of the
parton which produces a 5-10 GeV particle is of order $10^{-1}$, and this is
outside the region where on the basis of the HERA data one expects the
effects of the CGC to be important.  Nevertheless, nuclei might be different
than protons, so it is not a priori impossible.

The crucial test of these two different mechanisms is the comparison
of dA scattering to pp.  If there is suppression of jet in dA collisions, 
then it is an initial state effect.  The experiments were performed, and all 
there is little initial state effect for centrally produced 
jets.\cite{jetunsupp}
The suppression of centrally produced jets in AA collisions at RHIC is indeed
due to final state interactions, that is jet quenching.

This is not in contradiction with the existence of a CGC.  The particles
which control the multiplicity distribution in the central region are
relatively
soft, and arise from $x \sim 10^{-2}$.  To probe such small $x$
degrees of freedom at high transverse momentum at RHIC requires 
that one go to the forward region.\cite{dumitru}-\cite{gelis}

\begin{figure}[thb]
\vspace*{0.5in}
    \begin{center}
        \includegraphics[width=0.350\textwidth]{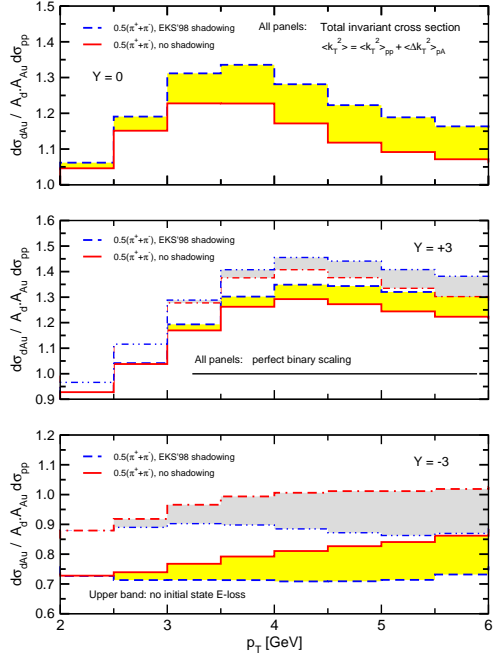}
\begin{minipage}[hbt]{6in}
        \caption{{\small The expectations of classical multiple
scattering for the rapidity and $p_T$ dependence (with and without shadowing) 
for the $p_T$ distribution in $D+Au$ collisions from {\protect\cite{Vitev:2003xu}}.
 }
\label{vit}}
\end{minipage}    \end{center}
\end{figure}

If one uses naive Glauber theory to compute the effects of shadowing
by multiple scattering,
one expects that if one goes into the forward region of the deuteron, the probe
propagates through more matter in the nucleus.  This is because we
probe all of the gluons with $x$ greater than the minimum $x$ of the nucleus
which can be seen by the deuteron.  Going more forward makes this minimum 
x smaller.  Now multiple scattering will produce more particles 
at some intermediate value of $p_T$.  (At very high $p_T$, the
effects of multiple scattering will disappear.)
This is the source of
the Cronin peak and it is expected to occur at $p_T$ of $2-4~GeV$.  Clearly
the height of this peak should increase as one goes more forward on the side
of the deuteron, and should increase with the centrality of the 
collision.\cite{vitev}
A result of such a computation is shown in Fig. \ref{vit}.

\begin{figure}[thb]
    \begin{center}
        \includegraphics[width=0.50\textwidth]{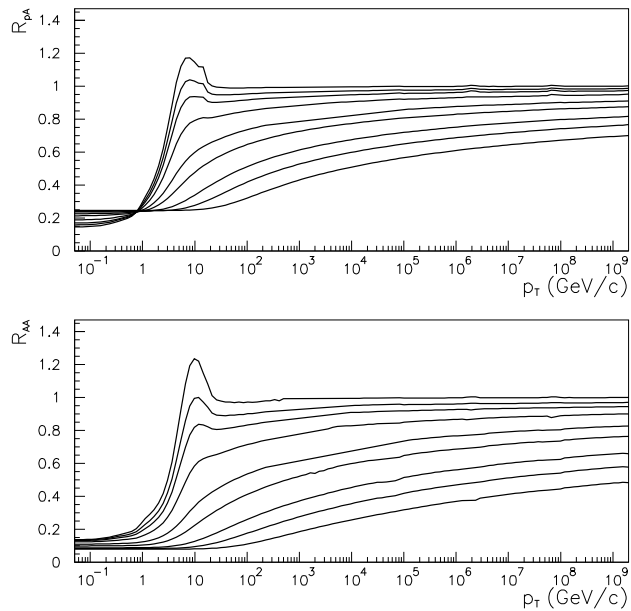}
\begin{minipage}[hbt]{6in}
        \caption{{\small The gluon intrinsic gluon distribution function as a
function of $p_T$ for different pseudo-rapidities{\protect\cite{kw2}}. 
 }
\label{kweid}} \end{minipage} 
    \end{center}
\end{figure}

Classical rescattering effects are included in the computation of the 
properties
of the CGC.  There is another effect however and that is quantum evolution
generated by the renormalization group equations.  It was a surprise
that when one computed the evolution of the gluon distribution function 
including
both effects, the quantum evolution dominated.  This means that the height
of the Cronin peak, and the overall magnitude of the gluon distribution
decreased  as one went from backwards to forward angles.\cite{kkt}-
\cite{kw2}  The results of one such computation are shown in Fig. \ref{kweid}

It was also a surprise how rapid the effect set in.

\begin{figure}[htb]
    \centering
       \mbox{{\epsfig{figure=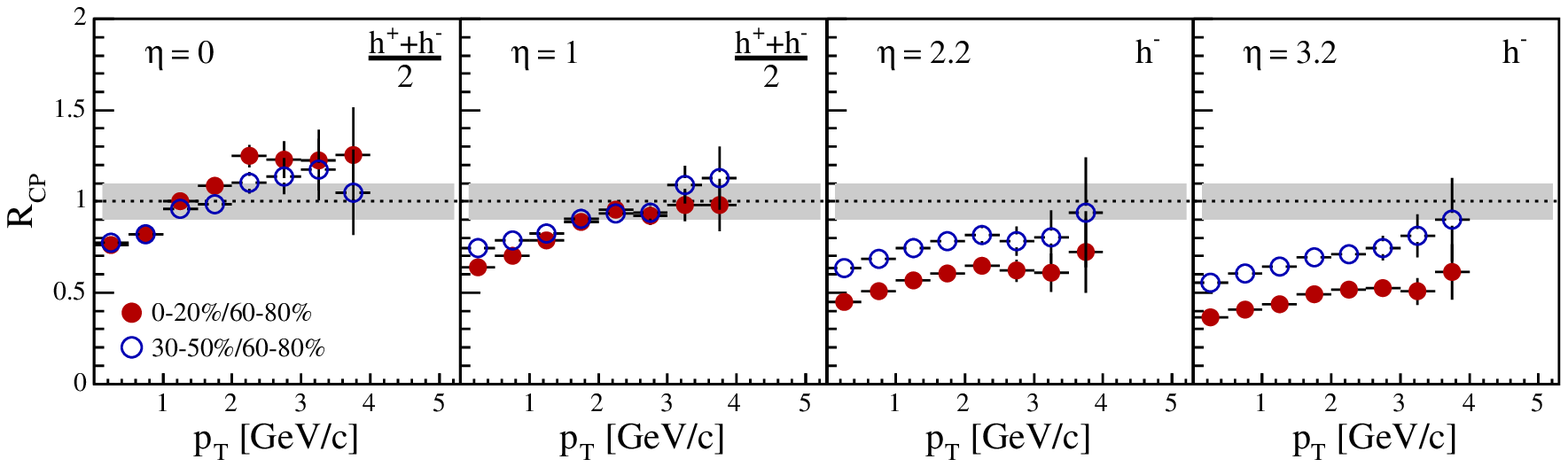,      
        width=1.0\textwidth,angle=0.}}}
\begin{minipage}[hbt]{6in}
        \caption{{\small The  central to peripheral charged and negative
hadron ratios $R_{CP}$ as a function of $p_T$
for various forward pseudorapidities in 200 AGeV $D+Au$ from 
BRAHMS\protect{\cite{Arsene:2004ux}}.
}
        \label{brahms}}\end{minipage}
\end{figure}

The Brahms experiment at RHIC recently presented data\cite{Arsene:2004ux} on the ratio
of central to peripheral transverse momentum 
distributions.
The ratio
$R_{CP}$ is defined in such a way that if the processes were due to
incoherent production of jets, then $R_{CP} = 1$.  A value less than one
indicates suppression, and a value larger than one indicates a Cronin 
type enhancement.  The results for a variety  of forward angles for $R_{CP}$
as a function of $p_T$ is shown in Fig. \ref{brahms} a.  There is clearly
a decrease in $R_{CP}$ as one goes to forward angles, in 
distinction from the predictions of classical multiple scattering.  
The effect is very rapid in rapidity, as was expected from computations 
of the CGC.
In Fig. \ref{brahms}, the ratio $R_{CP}$ is shown as a function of
$p_T$ for the forward pseudorapidity  $\eta \sim 3$ for less central and
more central events.  The ratio decreases for more central collisions,
against the expectation of classical multiple scattering and consistent with
the CGC hypothesis.

 Preliminary data\cite{starda}-\cite{phenixda} from all four experiments
on the rapidity dependence of the transverse distributions
in $D+Au$ suggest striking effects consistent with CGC.
This is a very active area of research both theoretically and experimentally
at this time.

This data suggest a Cronin enhancement on the gold side and the
depletion on the deuteron side, and as well a {definite} dependence on
centrality.  {When these $D+Au$ data become finalized, they could
  prove} that classical multiple scattering dominates on the gold side
{(large $x>0.01$)}, {but} quantum evolution {- i.e., deep gluon
  shadowing-} on the deuteron side {(small $x<0.01$)}.

By a ``happy coincidence'', these effects nearly cancel in the
mid-rapidity region, making RHIC {well suited} for studying QGP
effects at midrapidity for hard probes.

\subsection{The Developing Case for the CGC}

In addition to the results described above, the Color Glass Condensate will
be the subject of further experimental study at RHIC, LHC and eRHIC.
At RHIC, one can study forward backward correlations in
the forward direction in analogy with what was done for centrally produced
jets.  At LHC, in the forward region one measures relatively
large $p_T$ jets at $x \sim 10^{-6}$.  This provides a direct measurement
of the very small x gluon distribution function.  Eventually eRHIC would be
required to 
provide precision measurements of quark and gluon distribution functions
at small x in a variety of nuclei.

{\bf  The Color Glass Condensate hypothesis describes remarkably well generic
features of $ep$ measurements of properties of protons at small x.
It also successfully predicted the { the previously unexpected slow
growth of} multiplicity of produced particles with $\surd s$ and centrality
at RHIC.  The data from Brahms on the forward particle production
appear to be qualitatively  in accord with prediction 
of the CGC, and
the preliminary data from Phobos, Star and RHIC on this subject await 
submission for publication.  While the CGC hypothesis successfully
describes the data from disparate experimental measurements, it can
be further tested in a variety of new environments. }

\section{Conclusions}

Our criteria for the discovery of the Quark Gluon Plasma at RHIC are:

\begin{itemize}  

\item{\bf Matter at energy densities so large that the simple degrees of 
freedom are quarks and gluons. This energy density is that predicted by lattice
gauge theory for the existence of a QGP in thermal systems, and is about
$2 ~GeV/fm^3$}

\item{\bf The matter must be to a good approximation thermalized.}

\item{\bf The properties of the matter associated with the matter while
it is hot and dense must follow from QCD computations based on hydrodynamics,
lattice gauge theory results, and perturbative QCD for hard processes such
as jets}

\end{itemize}

All of the above are satisfied from the published data at RHIC.  A surprise
is the degree to which the computations based on ideal fluid hydrodynamics 
agree so well with elliptic flow data. This leads us to conclude that the 
matter produced at RHIC is a strongly coupled Quark Gluon Plasma (sQGP)
contrary to original expectations that were based on weakly coupled plasma estimates.

The case for the Color Glass Condensate is rapidly evolving into a compelling
case.  Much of the exciting new data from RHIC presented at QM2004 has 
yet to be published.  Nevertheless, the data from HERA taken together with
the data on particle multiplicities, and the data submitted for 
publication by Brahms make a strong case, which may become compelling with 
further reinforcement from results from the other experiments at RHIC, and 
future experimental tests at LHC and eRHIC.  This area continues to evolve rapidly
both experimentally and theoretically.

Although in our opinion, the case for the sQGP at RHIC is now
overwhelming, there are of course many important scientific issues not
yet addressed in the first three years of data.
 The experiments
 have demonstrated that a new form of matter, the sQGP, exists.
The harder long term task of mapping out more of its novel properties can now 
confidently proceed at RHIC.

\section{Acknowledgments}
This work was supported by the Director, Office of Energy
Research, Office of High Energy and Nuclear Physics,
Division of Nuclear Physics,
of the U.S. Department of Energy
under Contracts DE-FG-02-93ER-40764 and DE-AC02-98H10886.

\end{document}